\documentclass[journal]{IEEEtran}
\usepackage{amsmath}
\usepackage{amsfonts}
\usepackage{amssymb}
\usepackage[ruled]{algorithm2e}
\usepackage{array}
\usepackage{tabularx}
\usepackage{booktabs}
\usepackage[caption=false,font=footnotesize,labelfont=rm,textfont=rm]{subfig}
\usepackage{textcomp}
\usepackage{stfloats}
\usepackage{url}
\usepackage{verbatim}
\usepackage{graphicx}
\usepackage{cite}
\usepackage{ragged2e}
\usepackage{multirow}
\usepackage{tabularx,booktabs,array}

\usepackage[colorlinks=true,
            linkcolor=blue,
            urlcolor=blue,
            citecolor=blue]{hyperref}

\IEEEaftertitletext{\vspace{-2\baselineskip}}

\begin{document}

\title{Mip-NeWRF: Enhanced Wireless Radiance Field with Hybrid Encoding for Channel Prediction
}
\author{Yulin Fu, Jiancun Fan,~\IEEEmembership{Senior Member,~IEEE}, Shiyu Zhai, Zhibo Duan, and Jie Luo
\thanks{
This work was supported by the National Natural Science Foundation of China under Grants No. 62471381. \textit{(Corresponding author: Jiancun Fan.)}

Yulin Fu, Jiancun Fan, Shiyu Zhai, Zhibo Duan, and Jie Luo are with the School of Information and Communications Engineering, Xi’an Jiaotong University, Xi’an, Shaanxi 710049, P.R. China (e-mail: fanjc0114@gmail.com).
}
}



\maketitle

\begin{abstract}
Recent work on wireless radiance fields represents a promising deep learning approach for channel prediction, however, in complex environments these methods still exhibit limited robustness, slow convergence, and modest accuracy due to insufficiently refined modeling. To address this issue, we propose Mip-NeWRF, a physics-informed neural framework for accurate indoor channel prediction based on sparse channel measurements. The framework operates in a ray-based pipeline with coarse-to-fine importance sampling: frustum samples are encoded, processed by a shared multilayer perceptron (MLP), and the outputs are synthesized into the channel frequency response (CFR). Prior to MLP input, Mip-NeWRF performs conical-frustum sampling and applies a scale-consistent hybrid positional encoding to each frustum. The scale-consistent normalization aligns positional encodings across scene scales, while the hybrid encoding supplies both scale-robust, low-frequency stability to accelerate convergence and fine spatial detail to improve accuracy. During training, a curriculum learning schedule is applied to stabilize and accelerate convergence of the shared MLP. During channel synthesis, the MLP outputs, including predicted virtual transmitter presence probabilities and amplitudes, are combined with modeled pathloss and surface interaction attenuation to enhance physical fidelity and further improve accuracy. Simulation results demonstrate the effectiveness of the proposed approach: in typical scenarios, the normalized mean square error (NMSE) is reduced by 14.3 dB versus state-of-the-art baselines.

\end{abstract}


\section{Introduction}
\IEEEPARstart{I}{n} typical wireless propagation environments envisioned for sixth-generation (6G) systems, channels exhibit strong dynamics in the temporal, spatial, and frequency domains due to multipath propagation and shadowing \cite{bg1}. Accurate channel modeling and prediction can reduce pilot overhead and provide useful priors, thereby improving spectral efficiency and link reliability, which is particularly important in highly dynamic scenarios such as vehicular networks (V2X), unmanned aerial vehicle (UAV) communications, and large-scale internet-of-things (IoT) deployments \cite{bg2,bg3,bg_i1}. In emerging massive IoT and ambient IoT systems, a huge number of low-cost devices are expected to share limited wireless resources, making frequent channel sounding and acquaring increasingly expensive \cite{bg_i2,bg_i3,bg_i4}. Consequently, the ability to infer channel characteristics at arbitrary locations from sparse observations has become an important enabler for scalable and efficient 6G networks.

Traditional channel modeling approaches can be grouped into three main categories: probabilistic models, deterministic models, and hybrid models. Probabilistic models characterize channel behavior via statistical distributions, including path loss models, fading models, and cluster models\cite{bg5}. Such models are computationally simple and highly parameterizable, but their accuracy and generalization is limited and lack scene-specific interpretability. Deterministic models\cite{bg6, bg7}, such as ray tracing, overcome these limitations by relying on explicit geometric and electromagnetic propagation principles. These methods require prior knowledge of the environment and simulate wave-environment interactions to produce path-wise solutions, the resulting physically interpretable outputs, however, come at the cost of high computational complexity and strong dependence on accurate geometry and material descriptions, which hinders large scale deployment. Hybrid methods seek a compromise between physical interpretability and statistical generality, for example by combining ray tracing with statistical corrections\cite{bg8}. Although such designs can improve performance, they do not fundamentally eliminate the trade-offs to purely statistical/deterministic approaches.

\begin{figure}[!t]
\centering
\includegraphics[width=3.5in]{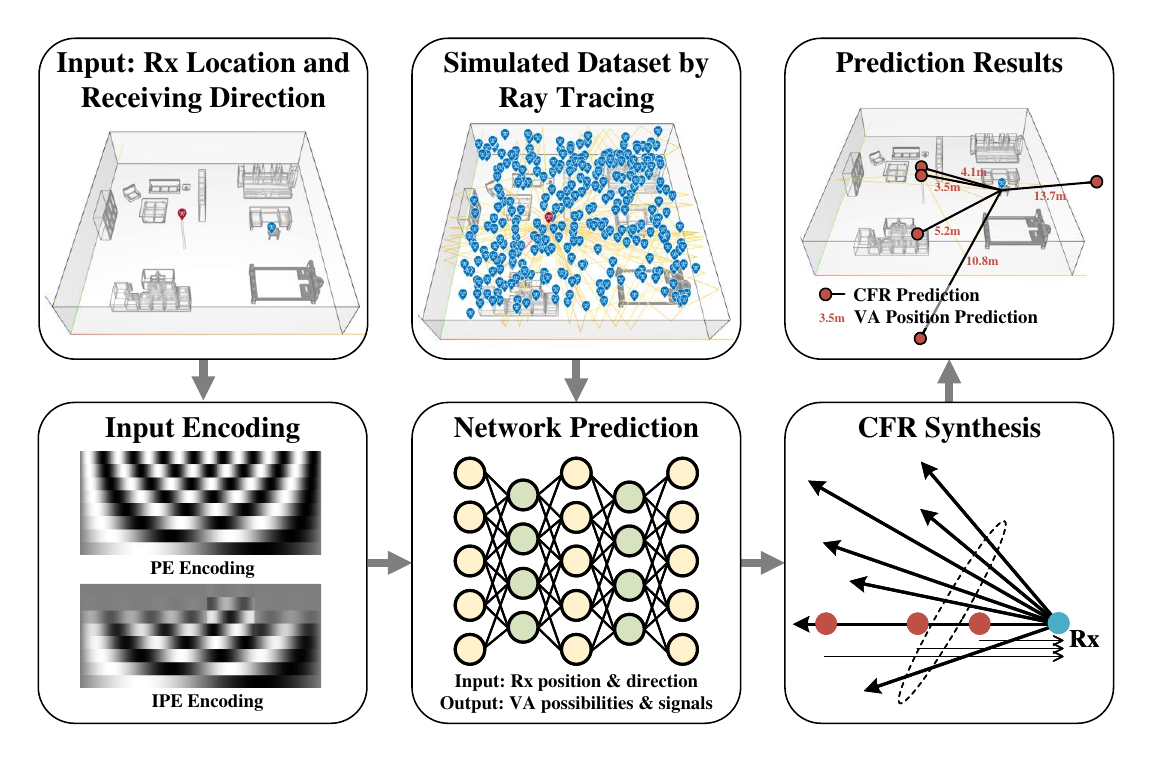}
\caption{Flowchart of Mip-NeWRF, which is trained to forecast CFR at any unknown receiver location.}
\label{fig_ill_5}
\end{figure}

\subsection{Related work}
Recent advances in deep learning have opened new avenues for channel prediction by learning the complex mapping between receiver locations and channel responses. These approaches can be broadly classified into three categories: direct network-channel prediction, neural ray tracing (neural RT), and wireless radiance field (WRF). Direct network-channel methods \cite{radiounet, streetnet, direct-net2, direct-net3, bg4} do not rely on explicit physical modeling and can automatically extract implicit environment-channel relationships from large datasets. For example, convolution neural network (CNN) based RadioUNet\cite{radiounet} predicts path loss distributions at arbitrary locations from environment maps; by fusing point cloud and building information and exploiting a convolutional autoencoder to extract spatial features, \cite{streetnet} achieves improved accuracy. However, direct network-channel approaches require dedicated acquisition and integration of environment data, and they typically suffer from poor interpretability and limited generalization.

By contrast, neural RT integrates ray-surface interaction mechanisms and geometric features into deep models \cite{winert, genert, nert-2, nert-3}. Neural RT reduces the computational burden of conventional ray tracing while retaining some physical interpretability. WiNeRT\cite{winert} pioneered this direction by using multilayer perceptrons (MLPs) to simulate ray-surface interactions along propagation paths. Subsequent works such as GeNeRT\cite{genert}, which incorporate relative geometric features and scatterer semantics, further improved accuracy and generalization. Nevertheless, neural RT still depends on detailed geometric or semantic scene priors and does not fundamentally resolve the sensitivity of ray tracing to geometry modeling errors.

WRF is inspired by Neural Radiance Field (NeRF)\cite{nerf} in computer science for image rendering, represent the spatial signal field implicitly and continuously with MLPs, can be trained directly from radio pilot measurements, and can be broadly categorized into omni-directional WRF and sparse-directional WRF. Omni-directional WRFs aim to reconstruct the radio field by predicting propagation characteristics over a dense set of spatial directions at each location, typically focusing on path-loss or power-related metrics. NeRF uses a MLP to learn continuous volumetric characterization, being able to distinguish the probability of surface existence and the intensity of the emitted light, and predicts images pixel by pixel by predicting the pixel values in each ray direction. By the analogy between optical rays and electromagnetic propagation, NeRF\textsuperscript{2} \cite{nerf2} migrated this paradigm to channel prediction: the network predicts attenuation and emission at sampling signal voxels, which are then weighted and aggregated into spatial spectrum. NeRA\cite{nera} extended this idea by incorporating environmental priors and skipping air voxels to accelerate inference. VoxelRF\cite{voxelrf} reduced network size and accelerated training by using a trilinear interpolated voxel grid representation. Although these approaches made progress, they remain time consuming. To accelerate synthesizing (note: \textit{render} indicating operations for visible light and \textit{synthesize} indicating operations for electromagnetic waves), WRF-GS\cite{wrf-gs} adopted explicit representations of virtual transmitters via 3D Gaussian splatting (3DGS) and synthesized channels by operating on the Gaussians along propagation paths rather than on all depth samples. This is also used in \cite{rf3dgs, 3dgs-2, 3dgs-3}, such as RF-3DGS\cite{rf3dgs} validated similar scheme in ISAC scenarios. However, Gaussian splatting inherently introduces smoothing that blurs resolution, depends on point cloud initialization, and requires maintaining a large number of Gaussians to preserve accuracy at scale. 

In contrast, sparse-directional WRFs, such as NeWRF \cite{newrf}, exploit a small number of dominant propagation directions as prior information and directly model complex channel responses along these directions, enabling more accurate phase-aware channel frequency response (CFR) prediction with substantially lower sampling overhead. This design significantly reduces sampling requirements while embedding propagation physics and improving interpretability, however, problems such as limited robustness to complex environments and scale variation (e.g., the prediction accuracy degrades rapidly as indoor volume increases), long training time, and slow convergence characteristic still exist.

\subsection{Contributions and Organization}
To address the drawbacks in existing sparse-directional WRF, we propose Mip-NeWRF, an enhanced physics-informed WRF for sparse indoor CFR prediction with task-specific coordinated improvements. The framework flowchart is shown in Fig.~\ref{fig_ill_5}. Positional encoding (PE) commonly used in WRF contains rich information but does not adhere to physical scale consistency, its high frequency components behave like noise and destabilize backpropagation, leading to poor cross-scale training consistency and low accuracy. Inspired by the anti-aliasing mechanism of Mip-NeRF\cite{mip-nerf}, we introduce a scale-consistent hybrid encoding that combines PE and integrated positional encoding (IPE) under an adaptive normalization framework, capturing the stable low-frequency spatial structures and fine-grained variational details simultaneously. Beyond encoding, Mip-NeWRF revisits the network architecture, optimization strategy, and the inherited optical rendering designs of existing WRF methods, adopts single shared network to accelerate convergence, and incorporates frequency-dependent electromagnetic priors in synthesis for reconstruction fidelity. Our main contributions are:

\begin{itemize}
    \item Scale-consistent hybrid encoding. We propose an adaptive normalization framework together with a PE--IPE hybrid encoding scheme. The proposed design maintains physical scale consistency and generalization ability across environments, and balances stable low-frequency representation with high-resolution spatial detail, exhibiting improved robustness to scene-scale variations and reduced prediction error.
    
    \item Shared network and stabilized training strategy. A single MLP serves both coarse and fine sampling stages, enabling effective information sharing while reducing model redundancy. Combined with curriculum learning, warm-up scheduling, and gradient clipping, the proposed framework accelerates and stabilizes training.
    
    \item Physics-aware channel synthesis. We incorporate frequency-dependent propagation priors into channel synthesis by explicitly compensating for path attenuation and interface interaction losses. Fresnel-based TE/TM modeling is adopted to account for polarization-dependent attenuation, improving accuracy while reducing learning burden.

    \item Extensive simulation validation. Mip-NeWRF is evaluated on representative scenarios and demonstrate significant gains: the normalized mean square error (NMSE) improves by 14.3 dB relative to NeWRF, and degrades only slightly as the scene scale increases. Effectiveness of each proposed module and cross-frequency generalization performance of the framework is further confirmed.
\end{itemize}

The remainder of this paper is organized as follows. Section~\ref{sec2} presents the system model and briefly reviews NeRF and NeWRF fundamentals. Section~\ref{sec3} details the Mip-NeWRF framework, including encoding, network, and synthesis modules. Section~\ref{sec4} describes the simulation setup and reports a comprehensive set of experiments validating the proposed approach. Finally, Section~\ref{sec5} concludes the paper.

\section{System Model}\label{sec2}
This section introduces the physical modeling foundations that are tightly integrated into the Mip-NeWRF framework. These models not only describe the underlying propagation mechanisms but also guide network learning and channel synthesis, forming the core of its physics-informed design.

\subsection{Wireless Channel Model}\label{sec2-1}
In typical wireless communication systems, a transmitted waveform experiences multiple forms of attenuation such as free-space path loss, reflection, transmission and diffraction, and the received signal is generally a superposition of multipath components. Assume the received waveform with $n$ multipaths is:
\begin{equation}\label{eq2-1}
    y(t) = \!\sum_{i=1}^{n}y_i(t) = \!\sum_{i=1}^{n}a_i x(t-\tau_i)
    =\! Ae^{j\varphi}\sum_{i=1}^{n}a_i s(t-\tau_i),
\end{equation}
where $s(t)$ is the origin baseband narrowband signal, $x(t)=Ae^{j\varphi}s(t)$ is the transmitted waveform, and $a_i$ and $\tau_i$ denote the complex attenuation coefficient and propagation delay of the $i$-th path, respectively.

Simultaneously perform continuous-time Fourier transforms (CTFT) on both sides of Eq.~\eqref{eq2-1}, we have:
\begin{equation}\label{eq2-2}
    H(f)\triangleq\frac{Y(f)}{X(f)}=\sum_{i=1}^na_ie^{-j2\pi f\tau_i},
\end{equation}
where $H(f)$ indicates the equivalent CFR, and $Y(f)$ and $S(f)$ are time-domain representation of $y(t)$ and $s(t)$, respectively. It is noted that the time-domain impulse response of such multipath channel is:
\begin{equation}\label{eq2-3}
    h(\tau) = \sum_{i=1}^n  a_i\delta(\tau-\tau_i),
\end{equation}
where $h(\tau)$ is exactly the time-domain counterpart of $H(f)$, exposes the discrete multipath components through their delays and complex gains.

Coefficient $a_i$ mainly consists of two parts, which are free-space propagation amplitude loss and interfaces interaction attenuation coefficient:
\begin{equation}\label{eq2-4}
    a_i = \!\!\!\!\!\underbrace{\bigg(\frac{c}{4\pi d_i f_c}\bigg)}_{\text{free-space propagation}} \!
    \underbrace{\bigg(     \prod_{j=1}^{n_r}\alpha_{i,j} \prod_{j=1}^{n_t}\beta_{i,j}
    \prod_{j=1}^{n_s}\delta_{i,j} \prod_{j=1}^{n_d}\eta_{i,j} \bigg)}_{\triangleq \zeta_i, \text{ }\text{ interfaces interaction attenuation}},
\end{equation}
where $d_i=\tau_ic$ is the propagation distance of the $i$-th path, $f_c$ is the carrier frequency, $n_r,n_t,n_s,\text{and }n_d$ are total interaction times of reflection, transmission, scattering and diffraction in the $i$-th path, and $\alpha_{i,j}, \beta_{i,j}, \delta_{i,j}, \text{and } \eta_{i,j}$ are attenuation coefficient of each interaction. Production of all interaction attenuation coefficient is denoted by $\zeta_i$ for short.

\begin{figure*}[!t]
\centering
\includegraphics[width=7.0in]{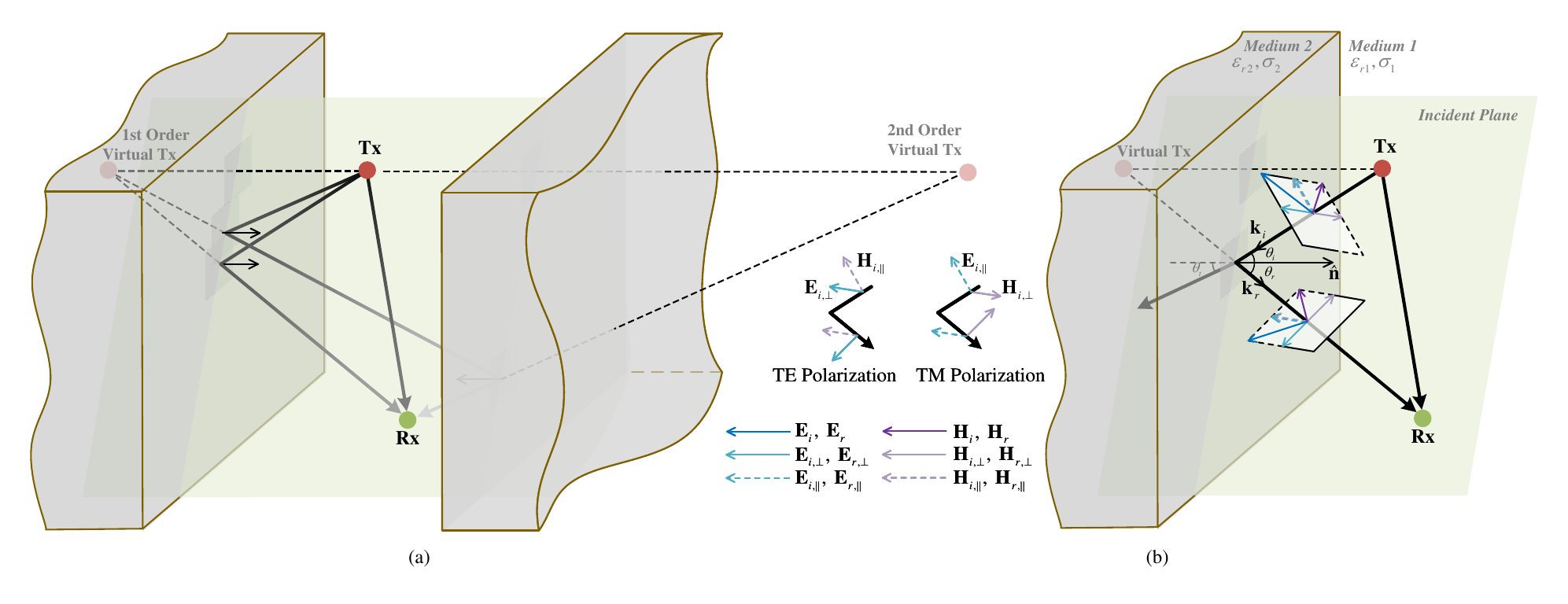}
\caption{Illustration of (a) wireless LoS and NLoS channel model, and (b) surface reflection model. $\mathbf{k}_i,\mathbf{k}_r,\hat{\mathbf{n}}$ are incident, reflection, and normal unit vector, respectively. $\mathbf{E}_i, \mathbf{E}_{i,\perp}, \mathbf{E}_{i,\parallel}$ / $\mathbf{H}_i, \mathbf{H}_{i,\parallel}, \mathbf{H}_{i,\perp}$ are synthetic, TE polarization, and TM polarization incident electric / magnetic field. The same applies to the reflection fields.}
\label{fig_ill_3}
\end{figure*}

In typical indoor wireless propagation scenarios, the strongest components consist of the line of sight (LoS) path and several specularly reflected NLoS paths \cite{pathmodel}, as illustrated in Fig.~\ref{fig_ill_3}(a). This indicates the transmission, scattering, and diffusion coefficients can be neglected to simplify analysis, i.e., omitting paths containting these components by setting $\beta_{i,j}\!=\! \delta_{i,j}\!=\! \eta_{i,j} \!=\!0$. Accordingly, $\zeta_i=\boldsymbol{1}_{\{n_t+n_s+n_d\}} \textstyle\prod_j\alpha_{i,j}$, where $\boldsymbol{1}_{\{x\}} $ returns 1 only when $x=0$ and 0 otherwise. The point located at distance $d_i$ from receiver along receive direction is treated as a virtual transmitter (also referred to as a virtual anchor, VA). A VA is therefore the mirror image of the transmitter with respect to the corresponding reflecting planes. The received signal can be regarded as emanating from these VAs and arriving at the receiver after free-space path loss and attenuation due to interactions at the reflecting interfaces. According to Eq.~\eqref{eq2-2} and Eq.~\eqref{eq2-4}, the CFR is:
\begin{equation}\label{eq2-5}
H = \sum_{i=1}^n  \zeta_i\frac{c}{4\pi d_i f_c}  e^\frac{-j2\pi f_c d_i}{c} ,
\end{equation}
in which $H$ contains all channel information, and is just the channel prediction target of this passage.

\subsection{Surface Interaction Model}

This subsection provides the specific calculation method for $\zeta_i$ in Eq.~\eqref{eq2-5}. The reflection and transmission behaviors of electromagnetic wave follows Fresnel's law. As shown in Fig.~\ref{fig_ill_3}(b), assume all materials are uniform, non-magnetic dielectrics ($\mu_r\!=\!1$) without birefringence or anisotropy, the synthetic electric and magnetic fields are denoted by $\mathbf{E}_i, \mathbf{E}_r$ and $\mathbf{H}_i, \mathbf{H}_r$, which can be decomposed into two orthogonal polarization components: transverse electric (TE) polarization (corresponding to $\mathbf{E}_{\perp},\mathbf{H}_{\parallel}$), and transverse magnetic (TM) polarization (corresponding to $\mathbf{E}_{\parallel},\mathbf{H}_{\perp}$), i.e.,
\begin{equation}\label{eq2-6}
\begin{alignedat}{2}
\mathbf{E}_i  = & \mathbf{E}_{i,\perp} + \mathbf{E}_{i,\parallel}
              & = & {E}_{i,\perp}\,\hat{\mathbf{e}}_{i,\perp} + {E}_{i,\parallel}\,\hat{\mathbf{e}}_{i,\parallel} \\
\mathbf{E}_r  = & \mathbf{E}_{r,\perp} + \mathbf{E}_{r,\parallel}
              & = & {E}_{r,\perp}\,\hat{\mathbf{e}}_{r,\perp} + {E}_{r,\parallel}\,\hat{\mathbf{e}}_{r,\parallel} \\
\mathbf{H}_i  = & \mathbf{H}_{i,\perp} + \mathbf{H}_{i,\parallel}
              & = & {H}_{i,\perp}\,\hat{\mathbf{e}}_{i,\perp} + {H}_{i,\parallel}\,\hat{\mathbf{e}}_{i,\parallel} \\
\mathbf{H}_r  = & \mathbf{H}_{r,\perp} + \mathbf{H}_{r,\parallel}
              & = & {H}_{r,\perp}\,\hat{\mathbf{e}}_{r,\perp} + {H}_{r,\parallel}\,\hat{\mathbf{e}}_{r,\parallel},
\end{alignedat}\end{equation}
where each $\hat{\mathbf{e}}$ is an unit orthogonal vector, satisfying:
\begin{equation}\begin{aligned}\label{eq2-7}
    &\hat{\mathbf{e}}_{i,\perp}\hat{\mathbf{e}}_{i,\parallel}
     =\hat{\mathbf{e}}_{i,\perp}\hat{\mathbf{k}}_i
     =\hat{\mathbf{e}}_{i,\parallel}\hat{\mathbf{k}}_i
     =0
    \\
    & \hat{\mathbf{e}}_{r,\perp}\mathbf{e}_{r,\parallel}
     =\hat{\mathbf{e}}_{r,\perp}\hat{\mathbf{k}}_r
     =\hat{\mathbf{e}}_{r,\parallel}\hat{\mathbf{k}}_r
     =0
    \\
    & \hat{\mathbf{e}}_{i,\perp}=\hat{\mathbf{k}}_i\times \hat{\mathbf{n}}
    \\
    & \hat{\mathbf{e}}_{i,\parallel}=\hat{\mathbf{e}}_{i,\perp} \times\hat{\mathbf{k}}_i,
\end{aligned}\end{equation}
where $\hat{\mathbf{k}}_i,\hat{\mathbf{k}}_r,\hat{\mathbf{n}}$ are incident, reflection, and normal unit vector, respectively.

Then the reflected waves can be represented by the product of incident waves and reflection coefficient as ${E}_{r,\perp} \!=\! r_{\perp}\!\times\!{E}_{r,\perp}$ and ${E}_{r,\parallel} \!=\! r_{\parallel}\!\times\!{E}_{r,\parallel}$, and $r_{\perp},r_{\parallel}$ can be calculated according to the Fresnel's law:
\begin{equation}\begin{alignedat}{2}\label{eq2-8}
r_{\perp}&=\frac{E_{\mathrm{r},\perp}}{E_{\mathrm{i},\perp}}&=&\frac{{\eta_1}\cos(\theta_i)-{\eta_2}\cos(\theta_t)}{{\eta_1}\cos(\theta_i)+{\eta_2}\cos(\theta_t)} 
\\
r_{\parallel}&=\frac{E_{\mathrm{r},\|}}{E_{\mathrm{i},\|}}&=&\frac{{\eta_2}\cos(\theta_i)-{\eta_1}\cos(\theta_t)}{{\eta_2}\cos(\theta_i)+{\eta_1}\cos(\theta_t)} ,
\end{alignedat}\end{equation}
where $\theta_i\!=\!\theta_r, \theta_t$ are angle of incident, reflection, and transmission, $\eta_1,\eta_2$ denote the intrinsic impedance of medium 1 (right in Fig.~\ref{fig_ill_3}(b)) and medium 2 (left), calculated by:
\begin{equation}\label{eq2-9}
    \eta = \sqrt{\frac{j\omega\mu}{\sigma+j\omega\varepsilon}},
\end{equation}
where $\omega\!=\!2\pi f$ is the angular frequency, $\varepsilon\!=\!\varepsilon_0\varepsilon_r$, $\mu\!=\!\mu_0\mu_r$, and $\sigma$ are permittivity, permeability, and conductivity, respectively. The variation of the complex reflection coefficient (and reflection power) of common materials with the incident angle is experimentally analyzed and summarized in subsection~\ref{sec-ref}.

Recall that $\zeta_i=\boldsymbol{1}_{\{n_t+n_s+n_d\}} \textstyle\prod_j\alpha_{i,j}$, but simply using $\alpha_{i,j}$ as a representation of one reflection is an approximate expression which ignores polarization mixing and rotation. In most cases, the reflected waves are irregular elliptically polarized waves. If the receiver has a polarization sensitivity unit vector $\mathbf{p}_r \!=\! \begin{bmatrix} p_{r,\perp}\!\!\! & p_{r,\parallel} \end{bmatrix} ^\top$, the complex reflection coefficient at a single reflection can be expressed as projection between incident and reflection waves:
\begin{equation}\begin{aligned}\label{eq2-10}
    \alpha_{i,1} = \mathbf{p}_r^H\mathbf{R}\mathbf{p}_i
                 = r_{\perp}p_{r,\perp}^* \frac{{E}_{i,\perp}}{\|\mathbf{E}_i\|}
                 +r_{\parallel}p_{r,\parallel}^* \frac{{E}_{i,\parallel}}{\|\mathbf{E}_i\|},
\end{aligned}\end{equation}
where $\mathbf{p}_i$ is unit vector of incident electric field intensity $\mathbf{E}_i$, and $\mathbf{R}\!=\! \text{diag} (r_{\perp},r_{\parallel})$ is the reflection operator.

For multiple reflections, the Jones matrix is introduced for precise representation:
\begin{equation}\begin{aligned}\label{eq2-11}
    \prod_j\alpha_{i,j} = \mathbf{p}_r^H \bigg(\prod_{j=1}^{n_r}\mathbf{R}_j\bigg) \mathbf{p}_i,
\end{aligned}\end{equation}
where $\mathbf{R}_j$ is the reflection operator in global base and can be calculated by:
\begin{equation}\begin{aligned}\label{eq2-11-2}
    \mathbf{R}_j= \mathbf Q_j^H \begin{bmatrix} r_{\perp}^{(j)} & 0 \\ 0 & r_{\parallel}^{(j)} \end{bmatrix} \mathbf Q_j,
\end{aligned}\end{equation}
where $\mathbf Q_j$ is unitary rotation matrix from local to global base, assisting mapping local TE/TM components to global ones.

Nevertheless, model in Eq.~\eqref{eq2-11} is overly detailed for practical channel prediction and can only be realized in ray tracing style environments. Since the exact polarization components and their phases are not available in actual prediction conditions, we account only for the amplitude attenuation introduced by reflection. The $j$-th reflection coefficient is expressed as:
\begin{equation}\begin{aligned}\label{eq2-12}
    \alpha_{i,j}={\sqrt{\omega_\perp\lvert r_{\perp}\rvert^2 + \omega_\parallel\lvert r_{\parallel}\rvert^2 }},
\end{aligned}\end{equation}
where $r_\perp,r_\parallel$ are TE/TM polarization reflection coefficients and $\omega_\perp,\omega_\parallel$ are their power weights (typically set $\omega_\perp\!=\!\omega_\parallel\!=\!0.5$). In practice, this reflection factor can be instantiated by using multipath SLAM \cite{ref-slam-1,ref-slam-2,ref-slam-3} to recover and associate continuous specular reflection paths for a moving receiver, from which incidence angles are obtained and, together with the materials’ electromagnetic parameters, used to compute $\alpha_{i,j}$ (see Fig.~\ref{fig_a}). A detailed description of this implementation is beyond the scope of the present paper and will be presented in a forthcoming publication.

\begin{figure*}[!t]
\centering
\includegraphics[width=7.0in]{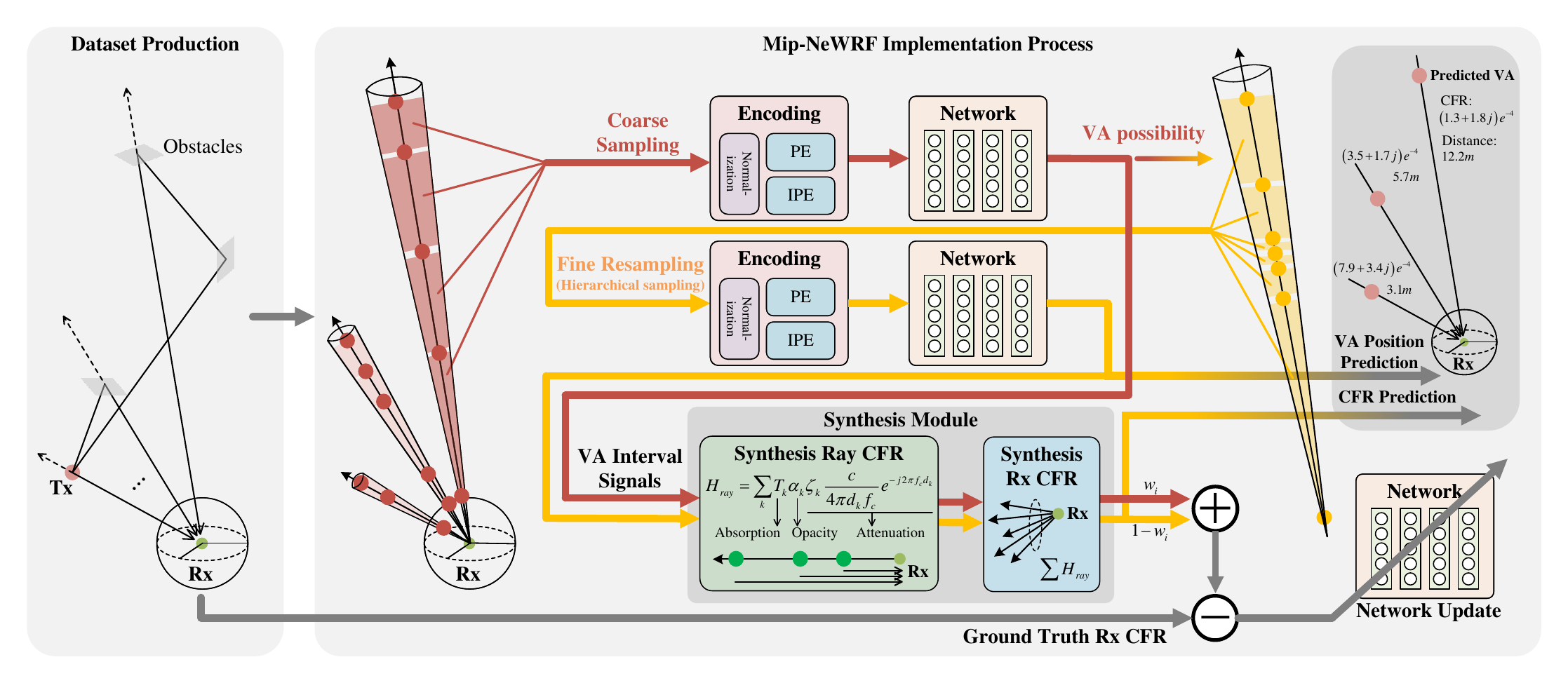}
\caption{Main implementation process of Mip-NeWRF. Two-stage sampling and prediction (coarse sampling in red and fine sampling in yellow) are adopted to enhance accuracy, where the fine sampling conducts importance sampling based on the VA distribution probability predicted in the coarse sampling. Two prediction stages share the same prediction network. The network output is the VA existence probabilities density and signal strengths at each sampling interval positions. These outputs are first synthesized to form the CFR of each ray, and then summed to obtain the CFR at the receiver position. During the training process, this value will be compared with the true value to update the network parameters.}
\label{fig_ill_2b}
\end{figure*}

\section{Mip-NeWRF Framework}\label{sec3}
Mip-NeWRF tries to provide accurate indoor CFR estimates with higher accuracy, faster convergence, and less affection by room scale. This section gives a comprehensive description of Mip-NeWRF, explains how the framework works, what hybrid encoding comprises of, how the MLP network is composed, and how the MLP output is synthesized into CFR.

\subsection{Framework Overview}
Mip-NeWRF framework implementation process is shown in Fig.~\ref{fig_ill_2b}. The framework consists of four parts, namely the sampling module, the encoding module, the network module (see Fig.~\ref{fig_ill_2a} for details) and the synthesis module. By inputting the position and viewing direction of the receiver, Mip-NeWRF can provide the CFR in that direction. 
The sampling module emits a ray along the input direction and performs interval sampling, then forwards the obtained samples to the encoding module. The encoding module transforms the sampled data and the ray direction into the representations consumed by the network. Encoding is the core of the radiance field because it determines how the physically meaningful inputs are presented to the model. The network predicts the VA probability and the (complex) signal amplitude for each sampling interval. The synthesis module propagates the interval-wise signals along the ray and composes the ray’s CFR at the receiver; the CFR contributions of all rays are then summed to produce the receiver CFR (see Eq.~\eqref{eq2-5}). The resulting prediction is compared with the ground-truth CFR and used to update the network parameters. To improve training efficiency and reconstruction quality we adopt two-stage sampling (coarse and fine), corresponding to stratified and hierarchical sampling in NeRF \cite{nerf}. After the coarse-stage VA probabilities are produced by the network, the fine-stage performs importance sampling using those probabilities as a prior, thereby concentrating samples near likely VAs. Both training and inference follow the same pipeline; the only difference is that during training the network parameters are updated from the prediction error. The following subsections describe the implementation and operation of each module in detail.

\subsection{Sampling}\label{sampling}

\begin{figure*}[!t]
\centering
\includegraphics[width=7.0in]{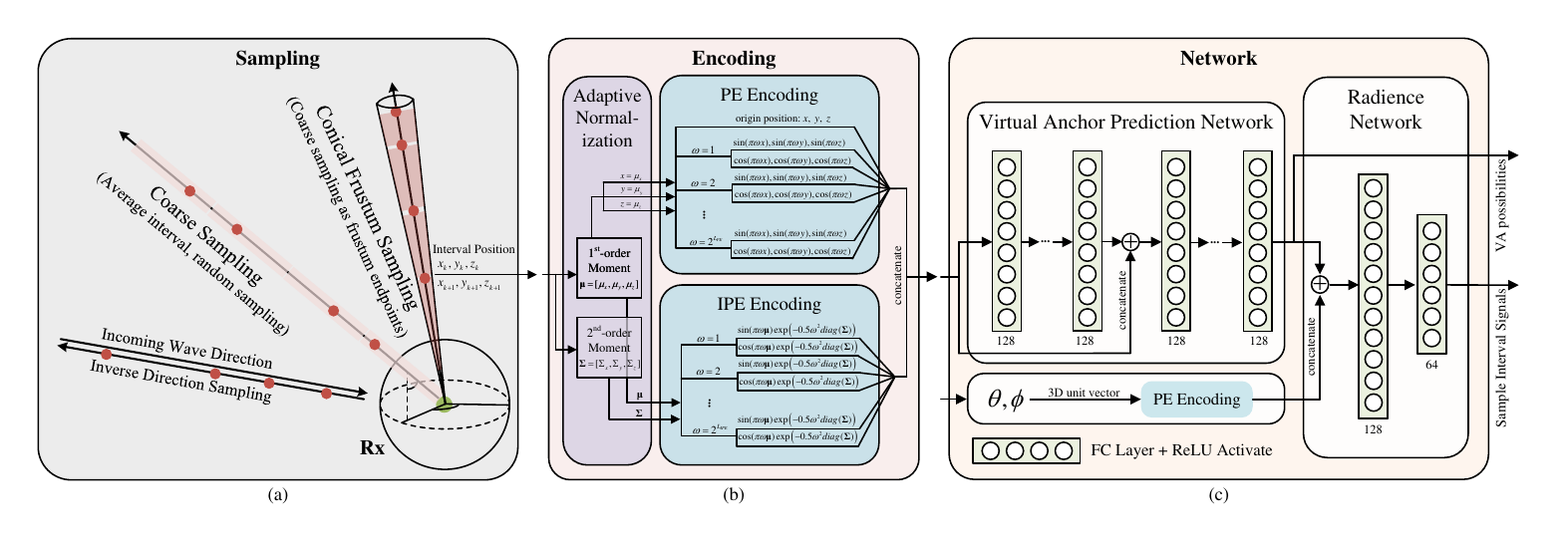}
\caption{The (a) sampling module, (b) encoding module and (c) network module of Mip-NeWRF. Sampling is carried out along the target receiving direction firstly. Subsequently, joint encoding of PE--IPE is carried out for all sampling intervals, and the results are sent into the network. The network forecasts signal strengths of each sampling intervals, and the output will pass through the synthesis module to output predicted CFR (see Fig.~\ref{fig_ill_2b}).}
\label{fig_ill_2a}
\end{figure*}

Reflected paths actually transmitted by the transmitter and received by the receiver can be seen as line-of-sight (LoS) rays emitted from corresponding virtual anchors (VAs, see Fig.~\ref{fig_ill_3}(a)). In a NeRF-style pipeline, sampling along a receive direction amounts to casting a ray in the opposite direction and sampling points along that ray (see Fig.~\ref{fig_ill_2a}(a)), which directly matches the VA interpretation. In other words, after sampling, positional encoding and the network modules, the model should be able to infer the locations of the VAs.

Accurate sampling in the vicinity of a VA substantially improves the prediction quality. Without any prior, sampling can only be random; hitting the VA then requires a large number of samples, which is highly inefficient because only samples near the VA are informative. The coarse-to-fine sampling strategy mitigates this issue: a coarse sampling pass first captures the global structure, the network’s outputs are used to estimate the probability that each coarse sample corresponds to a VA, a probability density function (PDF) is fitted from these estimates, and fine samples are then drawn according to that PDF so that sampling points are concentrated near VAs.

\subsubsection{Coarse Sampling}

As shown in Fig.~\ref{fig_ill_2a}(a), in coarse sampling we want to sample $m$ random intervals to form conical frustums along the direction of arrival (DoA). This sampling ray direction is represented as $\mathbf{r}(t_k) \!=\! \mathbf{o}+t_k\hat{\mathbf{d}}$, where $\mathbf{o}$ is receiver position, $\hat{\mathbf{d}}$ is unit inverse direction of incoming wave, and $t_k\!\in\![t_n,t_f]$ is sampling distance (also called depth along the ray). We firstly sample $m$ points from origin $\mathbf{o}$ along $\hat{\mathbf{d}}$ by uniformly partition $[t_n,t_f]$ into $m$ subintervals and perform one random point in each subinterval. This yields sample depths $t_2^c,\ldots,t_{m+1}^c$, and the ray-origin sample is fixed at $t_1^c\!=\!1e^{-3}$ (setting $t_1^c\!=\!0$ may cause singularities). These points form the endpoints of the sampling interval, i.e., the range of the $k$-th sampling interval is $[t_k^c,t_{k+1}^c]$.

\subsubsection{Fine Sampling}
Fine sampling follows an importance sampling strategy. Suppose the network has already produced a volume density (mass) $w_k$ (i.e., weight of VA existence, range of which is $[0,1]$) for each coarse interval $[t_k^c,t_{k+1}^c]$ after the coarse step. To make the probability distribution smoother, the weights are filtered first:
\begin{equation}\label{eq3-sampling-1}
w_k'=0.5\big( \max(w_{k-1},w_k)+ \max(w_{k},w_{k+1})\big),
\end{equation}

To facilitate resampling, it is assumed that the probability density is uniformly distributed in interval $[t_k^c,t_{k+1}^c]$ and its summation is mass $w_k'$, then the cumulative distribution function (CDF) along $\hat{\mathbf{d}}$ is defined as:
\begin{equation}\label{eq3-sampling-2}
F(t)=(t\!-\!t_1)\epsilon+\sum_{i=1}^{k-1} w_i'+\frac{t\!-\!t_k}{t_{k+1}\!-\!t_k}w_k',
\end{equation}
where $k$ is chosen to satisfy $i\!\in\![t_k^c,t_{k+1}^c]$, and $\epsilon$ is a base sampling density (set $\epsilon\!=\!0.01$ here) that ensures nonzero sampling probability in low-weight regions and thus prevents the sampler from collapsing onto incorrect VAs. We then draw $m$ uniform samples $u_j\!\sim\!\mathcal{U}(0,1), j\!=\!2,\ldots,m\!+\!1$ on $[0,1]$ and obtain the fine samples $t_k^f$ by inverse transform sampling, i.e. solving $F(t_k^f)\!=\!u_k$. The ray-origin sample is also fixed at $t_1^f\!=\!1e^{-3}$. These $m\!+\!1$ points also form the endpoints of the fine sampling conical frustums. This procedure concentrates samples in intervals with larger weights, yielding higher sampling density near likely VAs. In the sequel we do not distinguish between $t_k^c$ and $t_k^f$ unless explicitly point out, since they undergo the same downstream processing.

\subsubsection{Selection of Ray Direction}
The procedure for sampling along a given ray direction has been described above. We now discuss how to select the ray directions themselves. Each ray direction should be chosen as the inverse of a multipath DoA. Various approaches can be used for DoA estimation, including classical spectral estimation, Bayesian inference, and compressed sensing methods \cite{doa-estimate1, doa-estimate2, doa-estimate3}. In Mip-NeWRF, we assume that the estimated DoAs are known and modeled as the sum of the true DoAs and uniformly distributed noise $\vartheta\!\in\!\mathcal{U}(-\theta_e,\theta_e)$. In addition to these positive samples, a set of negative samples is also selected to balance the network input. The DoAs of these negative samples are randomly chosen, and their corresponding CFR labels are set to zero.

\subsection{Encoding}
\subsubsection{Scale-Consistent PE}\label{sec-pe}

NeWRF adopts the classical PE used in NeRF, expressed as $\gamma(\mathbf{x})\!=\![\gamma(x), \gamma(y), \gamma(z)]$, in which $\mathbf{x}\!=\!(x,y,z)$ is a sampling point and $\gamma$ is applied to three coordinate values separately, with $\gamma(x)$ expressed as: 
\begin{equation}\label{eq2-nerf-1}
    \gamma(x) \!\!=\!\! \left[
    \text{sin} (\!\pi x\!),  \text{cos} (\!\pi x\!),  \cdots,
    \text{sin} (\!2^{\!L\!-\!1}\!\pi x\!),  \text{cos} (\!2^{\!L\!-\!1}\!\pi x\!)         \right],
\end{equation}
where $\gamma$ is applied similarly for $y,z$, and $L$ is the encoding dimension. These results will be concatenated together and sent to the network. Eq.~\eqref{eq2-nerf-1} is definition of PE, implies that the encoding result depends solely on the position coordinates.

In the visual domain, images are typically normalized so that relative scene scale is approximately fixed. In wireless communications, however, spatial-scale variations directly affect electromagnetic phase and the geometry of reflection paths. Consequently, although the standard positional encoding carries useful information, it does not satisfy a physical scale-consistency constraint, and we observe a pronounced degradation of CFR prediction accuracy as room size increases.

To remedy this, we apply an adaptive normalization to each sampled coordinate $\mathbf{x}\!=\!(x_k,y_k,z_k)$. For the $x$-coordinate we perform:
\begin{equation}\label{eq3-1}
x_k' = \frac{x_k - x_{\min}}{2^{\left\lfloor \log_2(\mathrm{range}_x) \right\rfloor}},
\end{equation}
where $\mathrm{range}_x \!=\! x_{\max} \!-\! x_{\min}$ and $x_{\min},x_{\max}$ are the minimum and maximum $x$-values of the room range. The power-of-two normalization denominator is adopted solely to match the dyadic frequency hierarchy of NeRF style PE, and enables the normalized coordinate aligns with the frequency cascade used in PE across scenes of different absolute size. The positional encoding then becomes
\begin{equation}\label{eq3-2}\begin{aligned}
\gamma_s(x_k) = \big[  &\sin(2^0\pi x_k'),\ \cos(2^0\pi x_k'),\ \cdots,\ \\& \sin(2^{L_x-1}\pi x_k'),\ \cos(2^{L_x-1}\pi x_k') \big],
\end{aligned}
\end{equation}
with $L_x \!=\! 1 \!+\! \big\lceil \log_2(\mathrm{range}_x / d_{\min}) \big\rceil$,
where $d_{\min}$ denotes the target encoding resolution, i.e., the minimum spatial separation that can be uniquely represented by the highest-frequency component in PE, and a typical value is $d_{\min}\!=\!0.02m$. This ensures the uniformity of the numerical scale corresponding to the highest frequency in different scenarios. For $y$ and $z$-coordinate perform similar procedure of Eq.~\eqref{eq3-1} and Eq.~\eqref{eq3-2}. Then the scale-consistent PE is expressed as $\gamma_s(\mathbf{x})\!=\![\gamma_s(x_k),\gamma_s(y_k),\gamma_s(z_k)]$. This process enforces scale consistency so that the minimum resolvable feature is comparable across all coordinate axes.

\subsubsection{Scale-Consistent IPE}\label{sec-ipe}

\begin{figure}[!t]
\centering
\includegraphics[width=3.5in]{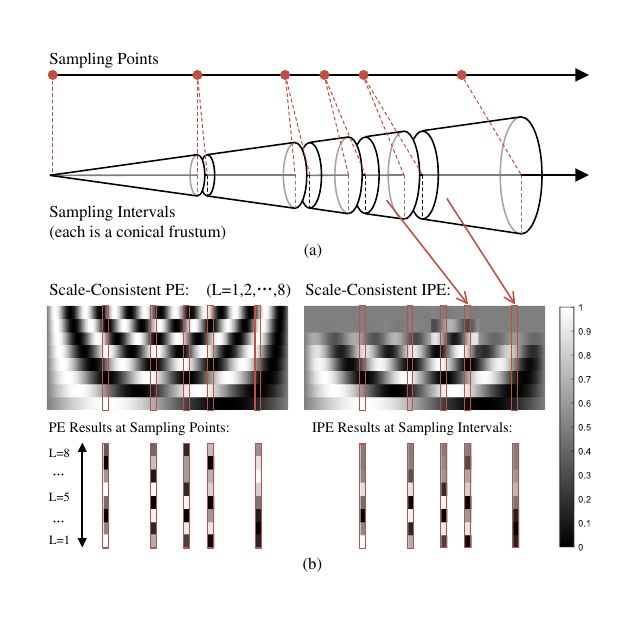}
\caption{(a) Illustration of sampling conical frustums and IPE. (b) Schematic diagram of comparison between scale-consistent PE and IPE. The low-pass effect of IPE becomes more obvious when the conical frustum is larger and the encoding frequency is higher.}
\label{fig_ill_6}
\end{figure}

Although the mentioned two-stage sampling strategy can roughly capture regions near VAs, it still depends on the coarse network’s preliminary importance estimates. Moreover, performing point sampling along an infinitesimally thin ray makes the encoding highly sensitive to small positional perturbations, causing large fluctuations and making the high-dimensional features behave like noise. Inspired by Mip-NeRF\cite{mip-nerf}, we introduce IPE to provide stability. IPE emits a conical frustum from the receiver along the target direction and performs stochastic interval sampling (already introduced in Subsection~\ref{sampling}). For each sampled frustum it computes the mean and covariance along the axis and by using these constructs a multivariate Gaussian as moment-matching approximate distribution of the original distribution, serving for encoding. As a result, IPE is inherently smoother and more robust to changes in sampling spacing and direction.

To explain IPE, recall from Subsection~\ref{sampling} that $m+1$ points $t_1,\cdots,t_{m+1}$ are sampled along $\hat{\mathbf{d}}$, corresponding to $m$ sampling intervals. Because IPE requires conical sampling, an important parameter is the cone aspect ratio (base radius divided by height), denoted by $\dot r$. A cone with parameter $\dot r$ is generated along the axial direction $\hat{\mathbf{d}}$. Cutting the sampling cone along the axis at depths $t_1$ to $t_{m+1}$ produces $m$ conical frustums, as shown in Fig.~\ref{fig_ill_6}(a). A point $\mathbf{x}\!=\!(x,y,z)$ belongs to the $k$-th frustum with axial bounds $[t_k,t_{k+1}]$ if and only if:
\begin{equation}\label{eq3-3}
\begin{aligned}
\Gamma(\mathbf{x},\mathbf{o},\hat{\mathbf{d}},\dot{r},t_k,t_{k+1})=\left(t_k<{\hat{\mathbf{d}}^{\mathrm{T}}(\mathbf{x}-\mathbf{o})}
<t_{k+1}\right)\\
\wedge
\left(\frac{\hat{\mathbf{d}}^{\mathrm{T}}(\mathbf{x}-\mathbf{o})}{\left\|\mathbf{x}-\mathbf{o}\right\|_{2}}>\frac{1}{\sqrt{1+\dot{r}^2}}\right)
,
\end{aligned}
\end{equation}
i.e., the point lies between the two axial planes and inside the cone opening defined by $\dot r$. Due to the DoA error, VA is assumed to occur at any position in the frustums with the same probability. Then for PE $\gamma(\cdot)$, the expected encoding over each frustum (i.e., the IPE) is given by the integral of $\gamma(\mathbf{x})$ with respect to the frustum’s spatial distribution. The IPE of the $k$-th frustum can be calculated by:
\begin{equation}\label{eq3-4}
\begin{aligned}
\mathbb{E}_{\text{origin}}[\gamma(\mathbf{x}_k)]=
\frac{\int\gamma(\mathbf{x})\Gamma(\mathbf{x},\mathbf{o},\hat{\mathbf{d}},\dot{r},t_k,t_{k+1})d\mathbf{x}}{\int\Gamma(\mathbf{x},\mathbf{o},\hat{\mathbf{d}},\dot{r},t_k,t_{k+1})d\mathbf{x}},
\end{aligned}
\end{equation}
where the integral is over the whole space.

Analytical integration of $\gamma(\cdot)$ over a conical frustum generally admits no closed-form solution, therefore, form a multivariate Gaussian $\mathcal{N}(\boldsymbol{\mu}_k,\boldsymbol{\Sigma}_k)$ by matching the frustum’s first and second moments, and use this Gaussian as a moment-matching approximation:
\begin{equation}\label{eq3-5}
\begin{aligned}
\mathbb{E}_{\text{origin}}[\gamma(\mathbf{x}_k)]&\approx
\mathbb{E}_{\mathbf{x}\sim\mathcal{N}(\boldsymbol{\mu}_k,\boldsymbol{\Sigma}_k)}[\gamma(\mathbf{x})].
\end{aligned}
\end{equation}

For a frequency level $l$, the positional encoding of the projected coordinate $\mathbf{k}\!^\top\!\mathbf{x}$, $\gamma(\mathbf{k}\!^\top\!\mathbf{x})\!=\![\sin(2^l\pi\mathbf{k}\!^\top\!\mathbf{x}), \cos(2^l\pi\mathbf{k}\!^\top\!\mathbf{x})]$, has the Gaussian-moment approximation:
\begin{equation}\label{eq3-6}
\begin{aligned}
\mathbb{E}_{\mathbf{x}\sim\mathcal{N}(\boldsymbol{\mu}_k,\boldsymbol{\Sigma}_k)}[\gamma(\mathbf{k}\!^\top\!\mathbf{x})] \!= &[\sin(2^l\pi\mathbf{k}\!^\top\!\boldsymbol{\mu})e^{-\frac{1}{2}(2^{l}\pi)^2\boldsymbol{\mathbf{k}\!^\top\!\Sigma\mathbf{k}}}, \\&\cos(2^l\pi\mathbf{k}\!^\top\!\boldsymbol{\mu})e^{-\frac{1}{2}(2^{l}\pi)^2\boldsymbol{\mathbf{k}\!^\top\!\Sigma\mathbf{k}}}].
\end{aligned}
\end{equation}

Because the positional encoding is formed by encoding each input dimension independently and concatenating the results, the expected encoding depends only on the marginal distributions of each dimension (i.e., the diagonal entries of $\boldsymbol\Sigma_k$) and not on cross-covariances. The basis vectors
$\mathbf{k}\!=\![1,0,0]^\top,[0,1,0]^\top,[0,0,1]^\top$ correspond to $x,y,z$, respectively. Considering scale consistency, the expected encoding at frequency level $l_x,l_y,l_z$ are:
\begin{equation}\label{eq3-7}
\begin{aligned}
&\mathbb{E}\big[\gamma_s(\mathbf{k}\!^\top\!\mathbf{x})\big]_{k,x},
\mathbb{E}\big[\gamma_s(\mathbf{k}\!^\top\!\mathbf{x})\big]_{k,y},
\mathbb{E}\big[\gamma_s(\mathbf{k}\!^\top\!\mathbf{x})\big]_{k,z}=
\\&
\big[
\sin(2^{l_x}\!\pi\mu_{k,x}')e^{-\!\tfrac{1}{2}\!(2^{l_x}\!\pi)^2\Sigma_{k,xx}'},
\cos(2^{l_x}\!\pi\mu_{k,x}')e^{-\!\tfrac{1}{2}\!(2^{l_x}\!\pi)^2\Sigma_{k,xx}'}
\big],
\\&
\big[
\sin(2^{l_y}\!\pi\mu_{k,y}')e^{-\!\tfrac{1}{2}\!(2^{l_y}\!\pi)^2\Sigma_{k,yy}'},
\cos(2^{l_y}\!\pi\mu_{k,y}')e^{-\!\tfrac{1}{2}\!(2^{l_y}\!\pi)^2\Sigma_{k,yy}'}
\big],
\\&
\big[
\sin(2^{l_z}\!\pi\mu_{k,z}')e^{-\!\tfrac{1}{2}\!(2^{l_z}\!\pi)^2\Sigma_{k,zz}'},
\cos(2^{l_z}\!\pi\mu_{k,z}')e^{-\!\tfrac{1}{2}\!(2^{l_z}\!\pi)^2\Sigma_{k,zz}'}
\big],
\end{aligned}
\end{equation}
in which:
\begin{equation}\label{eq3-8}
\begin{alignedat}{3}
\mu_{k,x}' & \!=\! q_x(\mu_{k,x} \!\!-\!\! x_{\min}),&
\Sigma_{k,xx}' & \!=\!  q_x^2\Sigma_{k,xx},&
q_x & \!=\!  2^{-\lfloor\log_2(\mathrm{range}_x)\rfloor} ,
\\
\mu_{k,y}' &\!=\! q_y(\mu_{k,y} \!\!-\!\! y_{\min}),&
\Sigma_{k,yy}' & \!=\!  q_y^2\Sigma_{k,yy},&
q_y & \!=\!  2^{-\lfloor\log_2(\mathrm{range}_y)\rfloor} ,
\\
\mu_{k,z}' &\!=\! q_z(\mu_{k,z} \!\!-\!\! z_{\min}),&  \
\Sigma_{k,zz}' & \!=\!  q_z^2\Sigma_{k,zz},& \
q_z & \!=\!  2^{-\lfloor\log_2(\mathrm{range}_z)\rfloor},
\end{alignedat}
\end{equation}
where $\mu_{k,x},\mu_{k,y},\mu_{k,z}$ are the per-dimension means of $\boldsymbol{\mu}_k$ and $\Sigma_{xx},\Sigma_{yy}, \Sigma_{zz}$ denote the corresponding diagonal entries of $\boldsymbol\Sigma_k$. The full IPE is obtained by concatenating per-dimensional IPE at all frequency levels. Note: frequency levels $l_x,l_y,l_z$ for dimension $x,y,z$ are not the same, satisfying $l_x\!=\!0,\cdots,L_x\!-\!1$, $l_y\!=\!0,\cdots,L_y\!-\!1$, and $l_z\!=\!0,\cdots,L_z\!-\!1$, respectively. In subsection~\ref{sec-pe} we've already discussed $L_x \!=\! 1 \!+\! \big\lceil \log_2(\mathrm{range}_x / d_{\min}) \big\rceil$, which is decided by the target encoding resolution.

Finally, let's deduce how to calculate the first moment $\boldsymbol{\mu}_k$ and the second moment $\boldsymbol{\Sigma}_k$ of a conical frustum. Recall that sampling points along the target direction are parameterized by:
\begin{equation}\label{eq3-9}
\mathbf{r}(t) \!=\! \mathbf{o}\!+\!t\hat{\mathbf{d}}
\end{equation}
where $t$ is the axial parameter and the cone radius at $t$ is $r(t)\!=\!\dot rt$. The $k$-th sampling interval corresponds to $t\!\in\![t_k,t_{k+1}]$. Because the frustum is radially symmetric, the two radial moments are equal and we do not distinguish them. The frustum’s first- and second-order moments can be obtained by integrating along the axial coordinate.

The normalization constant for the axial integral is:
\begin{equation}\label{eq3-10}
\begin{aligned}
Z=\int_{t_k}^{t_{k+1}}\!\pi t^2\, dt = \frac{\pi}{3}\left( {t_{k+1}^3-t_k^3}\right).
\end{aligned}
\end{equation}

The axial first and second moments are therefore:
\begin{equation}\label{eq3-11}
\begin{aligned}
\mathbb{E}[t]=\frac{1}{Z}\int_{t_k}^{t_{k+1}} \!t\cdot\pi t^2\, dt =\frac{3}{4} \frac{ {t_{k+1}^4-t_k^4}}{ {t_{k+1}^3-t_k^3}},
\end{aligned}
\end{equation}
\begin{equation}\label{eq3-12}
\begin{aligned}
\mathbb{E}[t^2]=\frac{1}{Z}\int_{t_k}^{t_{k+1}} \!t^2\cdot\pi t^2\, dt =\frac{3}{5} \frac{ {t_{k+1}^5-t_k^5}}{ {t_{k+1}^3-t_k^3}}.
\end{aligned}
\end{equation}

By symmetry the radial mean is zero, i.e., $\mathbb{E}[t]\!=\!0$, and the radial second moment at a fixed $t$ (for the thin disk at $t$) is:
\begin{equation}\label{eq3-13}
\begin{aligned}
\mathbb{E}[r^2|t]& \!=\!\frac{\int_{-\dot{r}t}^{\dot{r}t} x^2 \!\!\cdot\! 2\sqrt{\!({\dot{r}t})\!^2\!\!-\!\!x^2}  \, dx}{\int_{-\dot{r}t}^{\dot{r}t}2\sqrt{\!({\dot{r}t})\!^2\!\!-\!\!x^2}  \, dx}  \!=\!  \frac{\frac{1}{4}\pi\dot{r}^4t^4}{\pi\dot{r}^2t^2} \!=\!\frac{1}{4}\dot{r}^2t^2,
\end{aligned}
\end{equation}
averaging this over the axial interval yields:
\begin{equation}\label{eq3-14}
\begin{aligned}
\mathbb{E}[r^2]&=\frac{1}{Z}\int_{t_k}^{t_{k+1}} \! \mathbb{E}[r^2|t]  \cdot\pi t^2\, dt =\frac{3\dot{r}^2}{20} \frac{ {t_{k+1}^5-t_k^5}}{ {t_{k+1}^3-t_k^3}}.
\end{aligned}
\end{equation}

Hence the frustum’s axial and radial expectations and variances are:
\begin{equation}\label{eq3-15}
\begin{aligned}
{\mu}_{k,t}&=\mathbb{E}[t]=\frac{3}{4} \frac{ {t_{k+1}^4-t_k^4}}{ {t_{k+1}^3-t_k^3}},
\\
{\mu}_{k,r}&=0,
\\
{\Sigma}_{k,t}&=\mathbb{E}[t^2]\!-\!\mathbb{E}[t]^2=   \frac{3}{5}\frac{ {t_{k+1}^5 \!-\! t_k^5  } }{ {t_{k+1}^3 \!-\! t_k^3}} \!-\!  \frac{9}{16}\frac{( {t_{k+1}^4 \!-\! t_k^4 )^2 } }{ ({t_{k+1}^3 \!-\! t_k^3})^2},
\\
{\Sigma}_{k,r}&= \mathbb{E}[r^2] \!-\! \mathbb{E}[r]^2 =\frac{3\dot{r}^2}{20} \frac{ {t_{k+1}^5-t_k^5}}{ {t_{k+1}^3-t_k^3}}.
\end{aligned}
\end{equation}

Finally, assembling the axial and radial contributions in the global coordinate frame gives the frustum Gaussian approximation:
\begin{equation}\label{eq3-16}
\begin{aligned}
\boldsymbol{\mu}_k&=\mathbf{o}+{\mu}_{k,t}\hat{\mathbf{d}},
\\
\boldsymbol{\Sigma}_k&= {\Sigma}_{k,t}\hat{\mathbf{d}}\hat{\mathbf{d}}^\top
+ {\Sigma}_{k,r} ( \mathbf{I} - \hat{\mathbf{d}}\hat{\mathbf{d}}^\top ).
\end{aligned}
\end{equation}

As shown in Fig.~\ref{fig_ill_6}(b), IPE exhibits a more pronounced low-pass filtering effect at higher frequencies (comparing PE and IPE in figure with $L\!=\!8$) and for larger conical regions (comparing the encoded results of IPE with $L\!=\!7$). At fine sampling stage, smaller conical frustums (corresponding to regions with a higher probability of containing VAs) retain detailed features, while less important regions are smoothed out through stronger low-pass filtering.

\subsubsection{PE--IPE Hybrid Encoding}
We note that IPE was originally designed to mitigate aliasing and artifacts when rendering continuous, smooth surfaces. Because IPE performs a local spatial low-pass averaging, it tends to smooth out high-frequency, locally concentrated energy and therefore cannot faithfully represent sharply localized peaks. Our objective, however, is to predict discrete, spike-like VAs and their associated CFRs, which means accurate prediction cannot solely depend on smooth encoding itself.  Consequently, in Mip-NeWRF we adopt a scale consistent PE--IPE hybrid encoding (see Fig.~\ref{fig_ill_2a}(b)) that preserves PE’s ability to capture high-frequency detail while incorporating IPE’s scale-aware low-pass behavior; this hybrid produces the best overall performance. Since PE operates on individual spatial points, we use the mean position $\boldsymbol{\mu}_k$ of each conical frustum as the PE input, as shown in Eq.~\eqref{eq3-16}. In addition, to ensure the overall scale perception of the scene, the normalization of the original coordinates is also input into the network, denoted as $\gamma_o(\mathbf{x}_k)$:
\begin{equation}\label{eq3-a1}
\gamma_o(\mathbf{x}_k)=[\mu_{k,x}',\mu_{k,y}',\mu_{k,z}'].
\end{equation}

Consequently, the hybrid encoding procedure can be summarized as follows: after performing conical frustum sampling, the mean position and the conical frustum itself are encoded using original coordinates, PE and IPE according to Eq.~\eqref{eq3-a1}, Eq.~\eqref{eq3-2} and Eq.~\eqref{eq3-7}, respectively. These encoding results are then concatenated and fed into the MLP network. During training, this design enables faster convergence (with IPE providing stable, smooth low-frequency signals) and lower final error (as PE captures fine spatial details with strong representational capacity), which is particularly beneficial for sparse channel sampling and cross-scene generalization..

\subsubsection{Directional Encoding}
To fully exploit directional information, we apply sinusoidal PE to the direction vectors. We set $L\!=\!5$, corresponding to a fixed resolution, since the angular variation range remains consistent across different scenarios. The elevation and azimuth angles, $\theta$ and $\varphi$, are respectively encoded (as defined in Eq.~\eqref{eq2-nerf-1}), and the resulting directional encoding is denoted as $\gamma(\theta,\varphi)$.

\subsection{Network and Training}\label{net_arc}

\subsubsection{Network Architecture}

The network mainly consists of two MLPs with 8 and 2 layers, respectively, using ReLU as the activation function. Except for the last layer, which has 64 nodes, all other layers contain 128 nodes. To prevent gradient vanishing, the original network encoded input is concatenated with the output of the fourth layer and fed into the fifth layer. These two MLPs are responsible for predicting the probability of each sampling interval (referred to as the VA prediction network) and the signal intensity of each interval (referred to as the radiance network). The output of the VA prediction network is concatenated with the PE-encoded ray direction $\gamma(\theta,\varphi)$ as the input to the radiance network. The network can be represented as:
\begin{equation}\label{eq3-net-0}
    f_{\text{Mip-NeWRF}} \Big(\gamma_s(\mathbf{x}),\mathbb{E}\big(\gamma_s(\mathbf{x})\big),\gamma(\theta,\varphi);\mathbf\Theta \Big)\xrightarrow{}
    \big( \sigma_k, x _k\big),
\end{equation}
where $\gamma_s(\mathbf{x}), \mathbb{E}\big(\gamma_s(\mathbf{x})\big), \gamma(\theta,\varphi)$ are scale consistent PE, IPE, and directional encoding, respectively, and $\mathbf\Theta$ is the collection of network parameters. $\sigma_k$ is the predicted VA volume density (i.e., probability), and $x_k=A_ke^{j\varphi_k}$ denotes the equivalent complex signal value. The network predicts $x_k$ in terms of its real and imaginary components rather than amplitude $A_k$ and phase $\varphi_k$, this is because the phase value exhibits discontinuities, jumping from $2\pi$ back to $0$, which can lead to singularities. The detailed structure of the network is illustrated in Fig.~\ref{fig_ill_2a}(c).

\subsubsection{Training Strategy}

During each training iteration, a fixed number of receiver locations are randomly selected from the training pool. The number of selected receiver is set to 128, but this is not the batch size. For each selected receiver, all corresponding DoA directions and additional negative sample directions are included (typical number is 5-10; too many negative samples can bias the network toward outputting zeros). The actual batch size equals 128 multiplied by the number of sampled rays per receiver (typically 10--30), so the per-iteration batch size is not constant.

After the data are fed into the network, the predicted outputs are synthesized to produce the coarse-sampled channel $H_c$ and fine-sampled channel $H_f$. The MLP $\mathbf\Theta$ is trained by firstly calculating NMSE between the network synthesizing results $H(\mathbf{o})$ and the ground truth CFR $\hat{H}(\mathbf{o})$:
\begin{equation}\label{eq3-net-1}
\begin{aligned}
\mathcal{L}\big(H(\mathbf{o}),\hat{H}(\mathbf{o})\big)
= \frac{\sum \lVert \hat{H} - H \rVert^2}{\sum \lVert \hat{H} \rVert^2}
\end{aligned}
\end{equation}
where $\mathbf{o}$ is receiver location. The loss is computed as a weighted difference for backpropagation, achieved by minimizing:
\begin{equation}\label{eq3-net-2}
\begin{aligned}
\min_{\mathbf{\Theta}} \;&\mathcal{L}\big(H_c(\mathbf{o}),H_f(\mathbf{o}),\hat H(\mathbf{o})\big)\\
&=\sum_{\mathbf{X}\in\mathcal{R}} 
w^c\mathcal{L}(H_c,\hat H)+w^f\mathcal{L}(H_f,\hat H)
\end{aligned}
\end{equation}
where $w^c$ and $w^f$ are weighting coefficients satisfying $w^c\!+\!w^f\!=\!1$, $\mathcal{R}$ is set of all receiver locations in a batch. We set $w^c\!=\!0.1$ and $w^f\!=\!0.9$. 

Because the channel amplitudes vary significantly across different paths, NMSE provides a more balanced gradient for weak signals than the Mean Square Error (MSE), preventing them from being overwhelmed. We also experimented with using ${\sum |\hat H - H_c|^2 /|\hat H|^2}$ as the loss function, which enhances gradients for weak signals but was found to be more sensitive to noise, resulting in less stable gradient descent. Since NMSE typically spans several orders of magnitude, it is expressed in logarithmic form as $\mathcal{L}_{dB} \!=\! 10\log(\mathcal{L})$. For instance, $-10$ dB corresponds to a $10\%$ error while $-20$ dB corresponds to $1\%$. In the experiments, we use $\mathcal{L}_{dB}$ as representation of the fine sampling prediction NMSE to measure the prediction error.

Unlike NeWRF, which trains two separate networks for coarse and fine sampling, our method uses a single shared network for both stages. This strategy yields improvement in training speed without sacrificing accuracy. We consider the network converged when the average validation error falls below $-3$ dB and shows no improvement for 1,000 consecutive iterations.

We observed that training on very large and complex datasets can lead to poor optimization or stalled loss. To mitigate this, we adopt a simple curriculum-learning scheme by partitioning the training set into blocks of samples. Training starts using samples drawn from the first block; whenever the average validation error drops below $-10$ dB and at least 500 training iterations have passed since the last block was added, a new block is included in the training pool.

Network training uses the Adam optimizer and ReduceLROnPlateau learning-rate scheduler with patience equals to 3 and decay factor equals to 0.6. To ensure stability, a learning rate warm-up is adopted at the beginning 500 iterations of the training, and gradient clipping is used.

\begin{figure}[!t]
\centering
\includegraphics[width=3.5in]{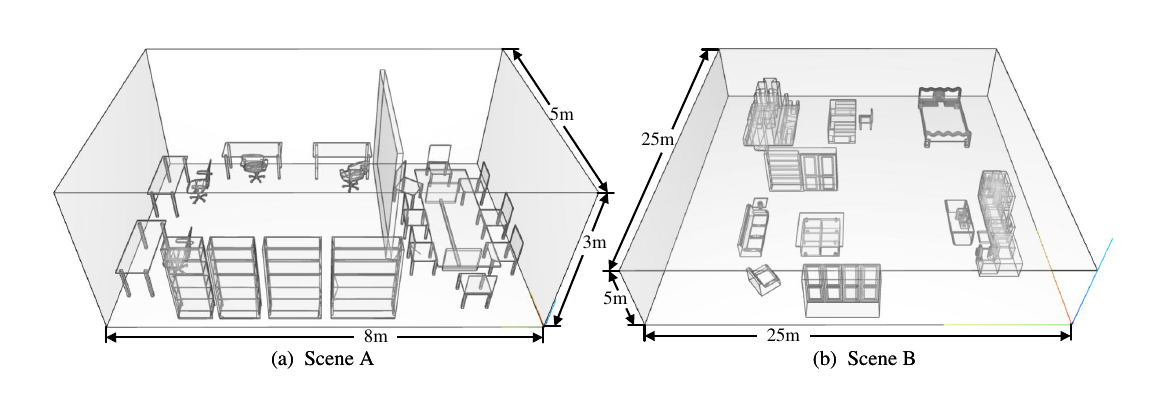}
\caption{Indoor scenes for ray tracing simulation. (a) Scene A, (b) Scene B.}
\label{fig_ill_7}
\end{figure}

\begin{figure}[!t]
\centering
\includegraphics[width=3.5in]{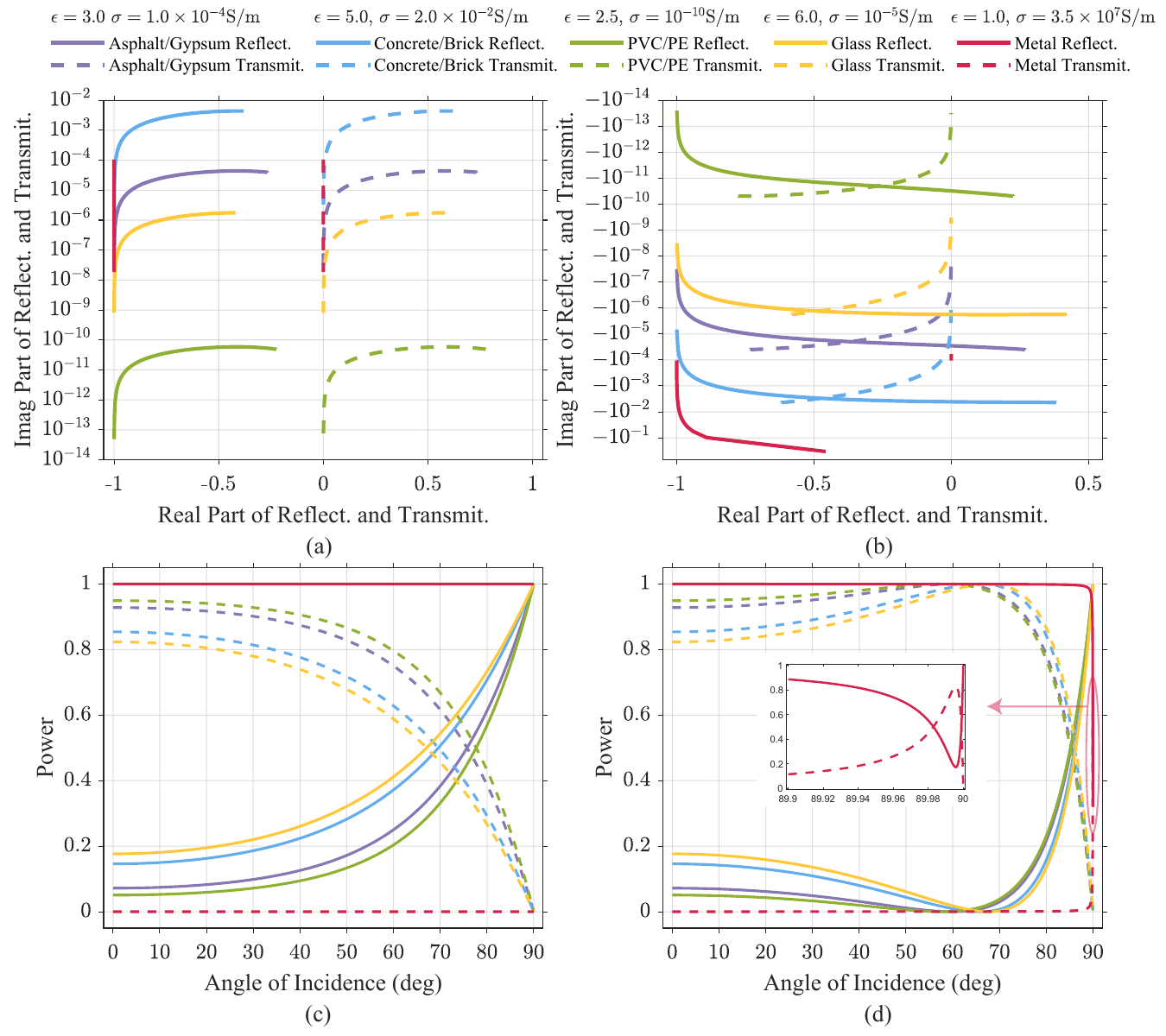}
\caption{Reflection and transmission characteristics of TE and TM polarized waves versus incidence angle. Complex reflection coefficient for (a) TE polarization, (b) TM polarization, and energy (power) variation for (c) TE polarization, (d) TM polarization are shown. Five dielectric materials are shown, with their parameters listed above the figure. Complex reflection coefficients are displayed in 2D plots. Solid lines denote reflected waves, dashed lines denote transmitted waves.}
\label{fig_a}
\end{figure}

\subsection{Synthesis}
The CFR is synthesized from the network outputs at each sampled location. First, note that for a unit-power transmitter the received signal at the receiver equals the channel. Accordingly, we predict the received signal when the transmitter transmits with unit power. Along a selected ray, the effective emission probability of the $k$-th sampling interval is computed from the network outputs as:
\begin{equation}\label{eq3-syn-1}
\nu_k=\underbrace{\Big(1\!-\!e^{-\sigma_k(t_{k+1}-t_k)}\Big)}_{\text{VA amplitude}} \quad\underbrace{\prod_{l=1}^{k-1} e^{-\sigma_l(t_{l+1}-t_l)}}_{\text{total transmittance}},
\end{equation}
in which $\nu_k$ is interpreted as the electromagnetic wave radiation amplitude strength. It is product of VA amplitude intensity (range of which is $[0,1]$, ) and residual intensity proportion reaches the receiver in the sense of volumetric rendering.

The receiver CFR is obtained by summing contributions from every path ray:
\begin{equation}\label{eq3-syn-2}
    H = \sum_{\text{rays}}\sum_{k=1}^m
    \frac{c}{4\pi \mu_{t,k} f_c}e^{-j\frac{2\pi \mu_{t,k} f_c }{c}}\nu_k\zeta_k{x}_k,
\end{equation}
where $\mu_{t,k}$ is the expected distance from the $k$-th frustum to the receiver, $\zeta_k$ denotes the interface interaction attenuation along the path from $\mu_{t,k}$ to the receiver (see Eq.~\eqref{eq2-4}), and ${x}_k$ is the VA transmit amplitude predicted by the network. Note that $\zeta_k$ only accounts for the amplitude attenuation caused by surface interactions, while the corresponding phase components are implicitly absorbed into the learned complex amplitude $x_k$. Factors $\nu_k$, $\zeta_k$, and free-space loss are not conflicting due to they act at different levels (probabilistic visibility and physical propagation mechanism) and are combined multiplicatively in the predicted result.

\section{Simulation Results}\label{sec4}
This section first presents the simulation setup, including simulation datasets and parameter configurations. Subsequently, the proposed method is compared with other baseline approaches, followed by ablation studies of the proposed modules along with corresponding analyses.

\begin{figure*}[!t]
\centering
\includegraphics[width=\textwidth]{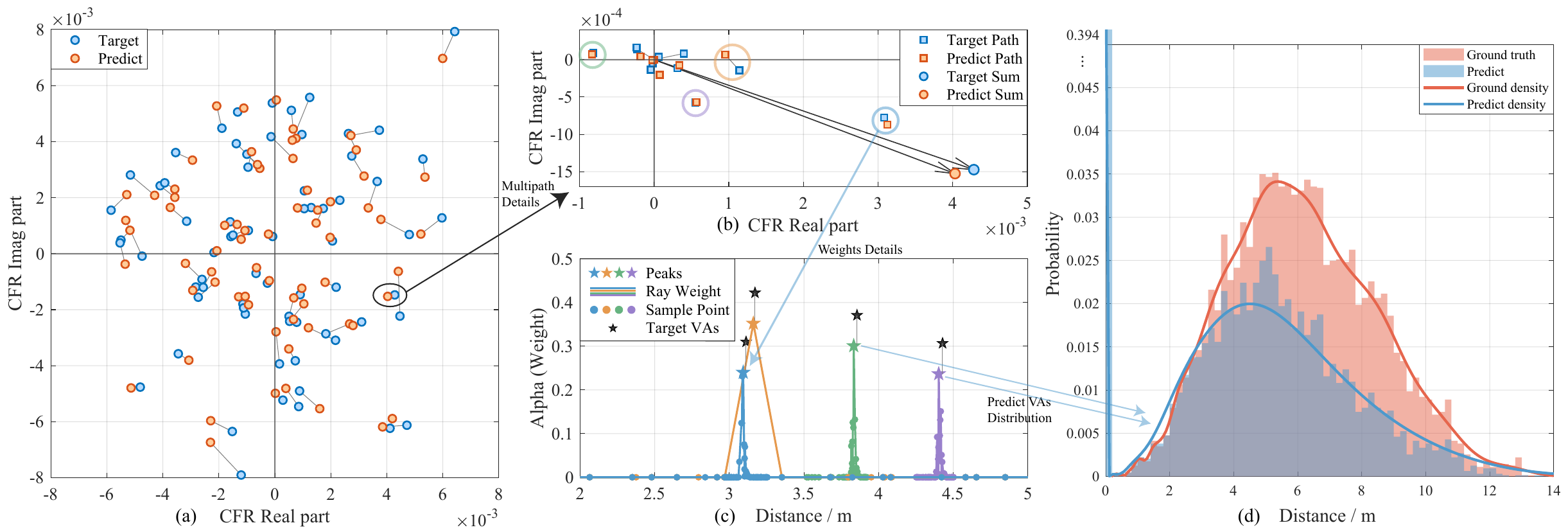}
\caption{Illustration of Mip-NeWRF results. (a) Real and imaginary (Re/Im) parts of simulated versus predicted CFR. (b) CFR decomposed by multipath and compared with VA positions. (c) Multipath ray-direction sampling weights (opacity indicates weight). (d) Distribution of peak ray sampling weights over all multipaths and receiver positions. Strong multipath signals generally yield VA predictions close to the receiver, while weak/long-distance paths sometimes result in zero outputs.}
\label{fig_e}
\end{figure*}

\subsection{Simulation Environment}
\subsubsection{Simulation Datasets}
The dataset was generated by the ray tracing simulator of MATLAB R2024a, employing the shooting and bouncing rays (SBR) method. Two scenes (can be found at https://github.com/Yulin-Fu/Mip-NeWRF-materials) were used (see Fig.~\ref{fig_ill_7}): Scene A matches the largest office room from NeWRF and measures $8{\times}5{\times}3$~m, while Scene B is an extra-large indoor environment of size $25{\times}25{\times}5$~m. For each scene we simulated channels at 3,000 (Scene A) and 6,000 (Scene B) receiver locations, respectively. Only paths with at most three interface interactions were included in the simulations. After removing the samples without receiving signals, there were 2,893 and 5,066 samples remaining.

\subsubsection{Parameter Configurations}

Mip-NeWRF is implemented in Python (Ubuntu 22.04) with PyTorch 1.13.1, training is performed on a machine equipped with an NVIDIA GeForce RTX 4090 and an Intel Core i7-14700K.

The ray tracing dataset is generated with gypsum material properties at a carrier frequency of $2.4\ \text{GHz}$ (Scene A) and $5.8\ \text{GHz}$ (Scene B). For sampling and encoding, the known DoA estimation error is set to $\theta_e \!=\!0.1^\circ$, the cone aspect ratio is set to $\dot{r}\!=\!0.0017$ (equal to $\sin \theta_e$, which is the angular resolution), and the target encoding resolution is $d_{\min}\!=\!0.02m$. Network architecture and main hyperparameters follow Subsection~\ref{net_arc}, and remaining experimental parameters are listed in Table~\ref{nn-params}.

\subsubsection{Reflection Coefficients}\label{sec-ref}
For common materials, the reflection coefficients and energy variations of TE and TM waves with respect to the incidence angle are shown in Fig.~\ref{fig_a}. Metallic surfaces tend to exhibit total reflection, while other materials show similar variation trends.

\begin{table}[htbp]
  \centering
  \caption{Mip-NeWRF Remaining Parameters}
  \label{nn-params}
  \small
  \begin{tabularx}{\linewidth}{@{}l *{2}{>{\centering\arraybackslash}X}@{}}
    \toprule
    Description & Scene A & Scene B \\
    \midrule
    Encoding dims & $10,10,9$ & $12,12,10$ \\
    Network input dims & $119$ & $139$ \\
    Sampling range & $[1e^{-3},15]$ & $[1e^{-3},30]$ \\
    Sampling number & $128$ & $256$ \\
    Negative ray number & $10$ & $5$ \\
    Learning rate & $1\!\times\!10^{-3}$ & $7.5\!\times\!10^{-4}$ \\
    Gradient clipping & $5\!\times\!10^{-3}$ & $5\!\times\!10^{-4}$ \\
    Block size & $3000$ & $2000$ \\
    \bottomrule
  \end{tabularx}
\end{table}

\begin{figure*}[!t]
\centering
\includegraphics[width=\textwidth]{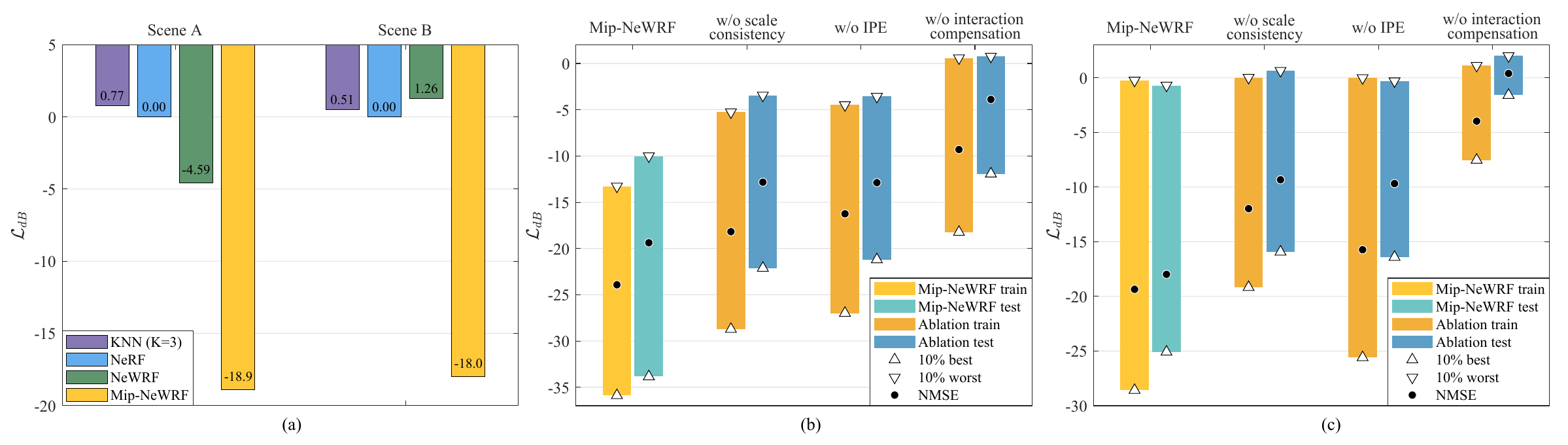}
\caption{Mip-NeWRF implementation results. (a) Average test set channel prediction NMSE among baseline methods including KNN, NeRF\textsuperscript{2}, NeWRF, and proposed Mip-NeWRF. (b) Ablation experiment results in scene A, including Mip-NeWRF, and Mip-NeWRF without scale-consistent encoding, IPE, or surface interaction compensation module. The upper/lower edges of the columns are the 10th/90th percentile of the error distribution. (c) Ablation experiment results in scene B.}
\label{fig_fb}
\end{figure*}


\begin{figure*}[!t]
\centering
\includegraphics[width=\textwidth]{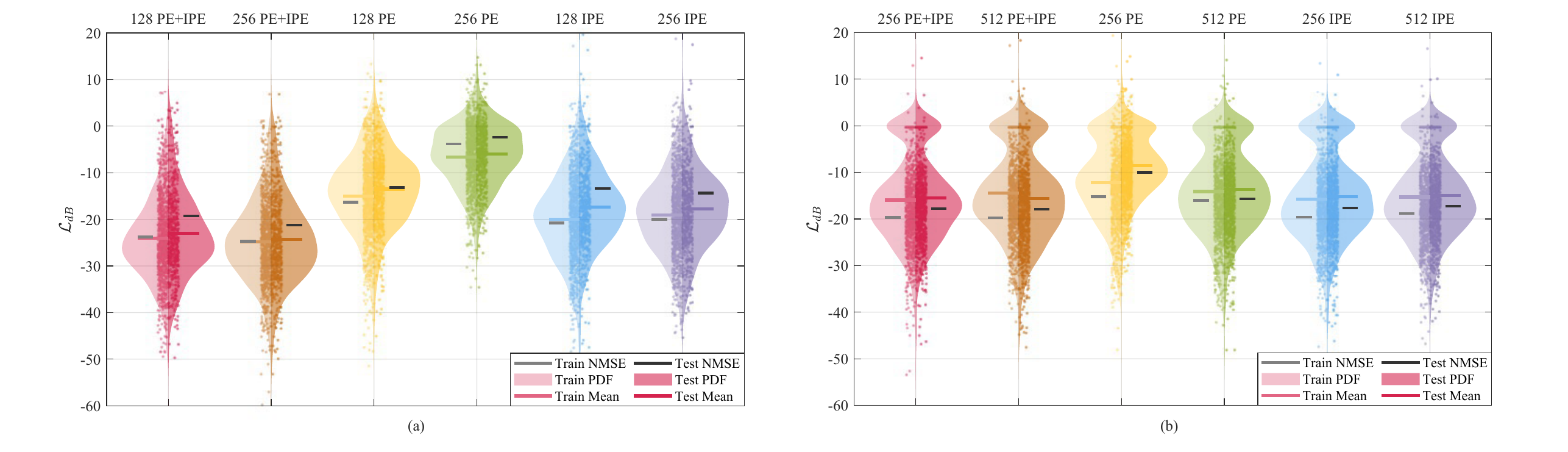}
\caption{Comparison of PE--IPE hybrid encoding versus PE-only and IPE-only. For Scene A we compare 128 and 256 intervals sampling-encoding; for Scene B we compare 256 and 512 intervals sampling-encoding. (a) Fitted error distribution in Scene A: left part corresponds to the training set, right to the test set. Scatter points represent the log-errors for individual receiver locations (a subset is plotted), the colored patches show the estimated probability density and the colored horizontal lines indicate the mean. Gray horizontal line marks NMSE. (b) Fitted error distribution in Scene B (same plotting conventions as (a)).}
\label{fig_f2}
\end{figure*}

\subsection{Mip-NeWRF Results}
A simple example in \textit{Scene A} is used to illustrate Mip-NeWRF’s predictions, as shown in Fig.~\ref{fig_e}(a). For any queried location the model outputs the predicted CFR, predictions are accurate at most locations, although errors occur at some receivers. Inspecting the multipath composition at a given receiver shows that Mip-NeWRF recovers channel components at the multipath level, and strong paths are predicted with high accuracy, as shown in subfigure (b). In subfigure (c), each predicted path corresponds to a sampled ray and an associated VA probability over the sampling interval, and the network localizes VAs reliably. Statistical analysis of VA distance detections in subfigure (d) indicates that nearby VAs are recovered consistently, while distant VAs, whose received amplitudes are weak, are often missed. Because negative samples are included during training, the network tends to output zero for very weak paths, effectively treating them as absent. In addition to NMSE, the detailed fitted error distribution of train/test samples can be found in Fig.~\ref{fig_f2}, used for more comprehensive evaluation.

Fig.~\ref{fig_fb}(a) illustrates the channel prediction NMSE of Mip-NeWRF and baseline methods in two scenarios, where a smaller $\mathcal{L}$ indicates better performance. The KNN method computes the target channel by weighting the channels of the $K\!=\!3$ nearest neighboring positions. However, it fails to effectively capture rapidly varying phase characteristics, resulting in poor performance. The NeRF\textsuperscript{2} network, as an example of omni-directional WRF, is trained on spatial spectrum generated from our sparsely measured channels, but due to presence of numerous negative samples, the network collapses to zero outputs. This indicates that the omni-directional WRF cannot achieve the goals under the sparse-directional WRF model. The prediction accuracy of NeWRF surpasses that of the first two methods in Scene A, yet in Scene B of larger sacle, its performance remains unsatisfactory even when the number of sampling points is increased to $768$ / $30m$. The proposed Mip-NeWRF achieves consistently better predictions in both environments, and its performance degrades only slightly as the scene scale increases. Note that the sparse-directional formulation is naturally aligned with the adopted specular reflection SBR simulator, therefore, the results primarily validate the effectiveness of Mip-NeWRF in geometry-dominant propagation conditions.

\subsubsection{Ablation Experiments}
We compare Mip-NeWRF with three ablated variants on Scene A and Scene B (see Fig.~\ref{fig_fb}(b) and (c)): (i) without the scale-consistent normalization, (ii) without IPE, and (iii) without interface-interaction attenuation compensation. Each ablation degrades performance by roughly 7 dB, 7 dB and 16 dB, respectively, demonstrating the effectiveness of the proposed components. Removing the scale-consistent normalization changes the effective resolution intervals of the positional encoding across scales implicitly, which harms generalization and makes the network prone to memorizing the training set. Omitting IPE removes the stable, low-frequency content supplied to the network and the remaining PE-dominated high-frequency components behave like noise and impede convergence. Finally, without explicit interface-interaction attenuation compensation the network is forced to implicitly learn these surface interaction effects, this extra learning burden grows with scene size and leads to rapid performance degradation. Note that the worst 10\% of cases in Scene B perform poorly, with NMSE approaching 0 dB. This is caused by certain receiver positions receiving very weak signals, for which the corresponding VAs are difficult to detect. In such cases the network fails to locate VAs across almost all multipath components and tends to output zero (see Fig.~\ref{fig_e}(d)), consequently the synthesized CFR is effectively zero. This behavior is consistent with the probability distribution shown in Fig.~\ref{fig_f2}(d) (discussed in next subsection).

\subsubsection{Hybrid Encoding Effectiveness}

Fig.~\ref{fig_f2} compares the test set error distribution of hybrid (PE--IPE) encoding with PE-only and IPE-only across two scenes. In training process, the hybrid encoding scheme achieves a faster takeoff (initial improvement) and consistently better validation NMSE. Examining the error distributions in (a) and (b), it reveals that Scene A exhibits an overall uniform NMSE distribution, whereas Scene B shows many points clustered at 0 dB. This clustering is caused by the greater complexity of VA distributions in the large scene, which leads the model to output zero for some receiver positions (a phenomenon consistent with Fig.~\ref{fig_e}(d)). Under the same number of samples the hybrid encoding yields much lower NMSE than PE-only or IPE-only; even when PE-only or IPE-only are given more samples to match the hybrid’s input dimensionality, the hybrid still retains an advantage, particularly in Scene A. This indicates that the improvement stems from increased representational capacity of the hybrid encoding rather than merely from larger input dimensionality.
\begin{figure}[!t]
\centering
\includegraphics[width=3.5in]{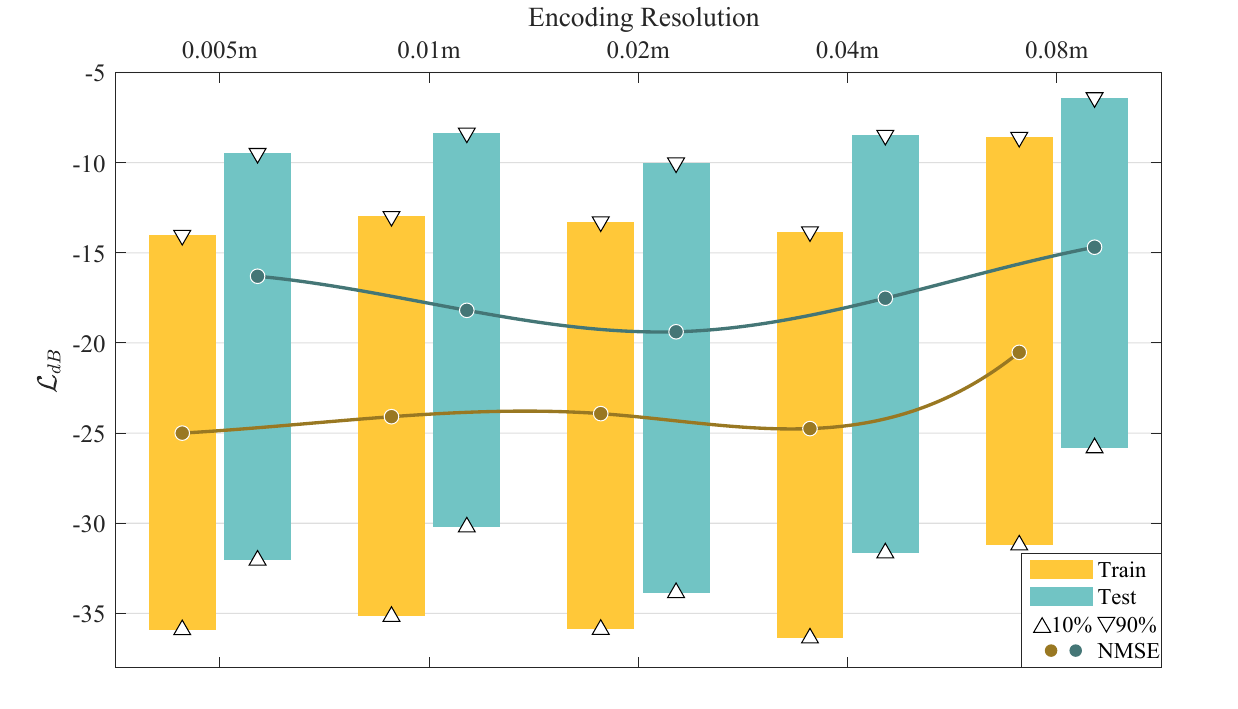}
\caption{Channel prediction NMSE error with different encoding resolution.}
\label{fig_f3}
\end{figure}

\vspace{-1em}
\subsubsection{Optimal Encoding Resolution}
Figure~\ref{fig_f3} shows the prediction NMSE in Scene A as a function of encoding resolution, the best performance is obtained at $d_{\min}\!=\!0.02m$. Low encoding resolution fails to provide sufficient informative content, whereas excessively high encoding resolution boosts the high frequency bands of the encoding, amplifying input discontinuities and thereby increasing noise. The encoding should be chosen to match the network’s effective receptive field so that the model can efficiently learn the relationship between position and VA characteristics.

\subsubsection{Impact of DoA Estimation Error}
Table~\ref{tab:doa-nmse} compares the impact of different DoA estimation errors on the prediction NMSE across Scene A dataset. The performance of both methods degrades as DoA error increases, while the proposed Mip-NeWRF consistently achieves higher prediction accuracy. This behavior is expected because sparse directional WRF frameworks rely on accurate DoA information to learn the distribution of VAs and infer the underlying propagation geometry. And the VA estimation error $\Delta x \!\approx\! d\tan\theta_e$, is directly related to the VA distance and DoA error. Consequently, the prediction performance is inherently sensitive to DoA errors, reflects a fundamental feature of geometry-aware channel modeling approaches rather than specific limitation of Mip-NeWRF. For an NMSE target of $-10$ dB, the required DoA accuracy is approximately $1^\circ$. Although such accuracy requirement is high, it is related to the array size, and can be satisfied by methods such as SBL \cite{doa-estimate1} under $16$-element uniform linear array (ULA).

\vspace{-1em}
\subsection{Other Experiments}
\subsubsection{Training Strategies}
Keeping the same encoding resolution in Scene A, when the VA prediction network has 6, 8 and 10 layers, the channel prediction NMSEs are $-17.16$ dB, $-18.89$ dB and $-19.25$ dB, respectively, with corresponding training speeds of $4.88$, $4.47$ and $4.12$ iterations/s. This indicates diminishing returns from increasing network capacity. With the VA prediction network fixed at 8 layers, using a single shared network vs. two separate networks yields NMSEs of $-18.89$ dB and $-19.07$ dB and training speeds of $4.46$ and $3.73$ iterations/s, showing that the single-network design substantially improves training throughput while only slightly affecting accuracy. Furthermore, for Scene B with 6,000 samples, omitting the curriculum-learning strategy degrades the prediction NMSE to $-13.34$ dB compared with $-17.98$ dB when the strategy is used, which we attribute to the increased task complexity preventing the model from learning a stable representation.

\begin{table}[!t]
  \centering
  \caption{NMSE under Different DoA Estimation Errors in Scene A}
  \label{tab:doa-nmse}
  \small
  \begin{tabularx}{\linewidth}{@{}l *{5}{>{\centering\arraybackslash}X}@{}}
    \toprule
    Method \textbackslash $\theta_e$ & $0.1^\circ$ & $0.2^\circ$ & $0.4^\circ$ & $0.8^\circ$ & $1.6^\circ$ \\
    \midrule
    NeWRF & $-4.59$ & $-4.21$ & $\mathit{-3.86}$ & $\mathit{-0.63}$ & $\mathit{0.40}$ \\
    Mip-NeWRF & $-18.94$ & $-16.63$ & $-13.57$ & $-11.19$ & $\mathit{-3.58}$ \\
    \bottomrule
  \end{tabularx}

  \vspace{0.2em}
  \begin{minipage}{\linewidth}
    \footnotesize
    \textit{Note:} Italic entries correspond to halved training learning rate.
  \end{minipage}
\end{table}

\subsubsection{Cross-materials and Cross-frequency Influences}
\begin{figure}[!t]
\centering
\includegraphics[width=3.5in]{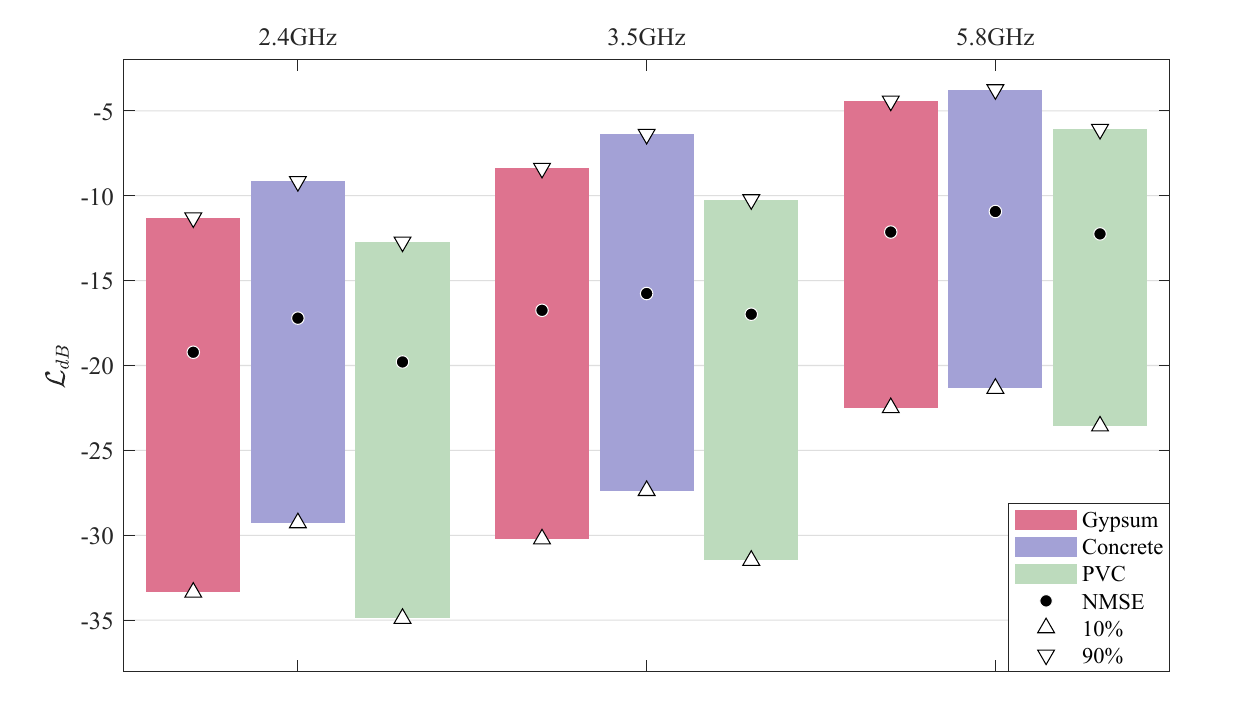}
\caption{Channel prediction NMSE error of the model trained on the $2.4$ GHz gypsum material ray tracing dataset when tested on datasets with different materials and frequencies. For each new dataset, a fine-tuning transfer training of 100 iterations was performed.}
\label{fig_f5}
\end{figure}

We compare the channel prediction NMSE of the proposed method under cross-material and cross-frequency conditions, as shown in Fig.~\ref{fig_f5}. It is noteworthy that although Mip-NeWRF is physics-informed, it still exhibits slight dependence on material properties and frequency. When material or frequency changes, the model performs poorly without transfer training, but with only 100 iterations of light fine-tuning, it achieves performance comparable to that on the original test set. Compared to normal training, fine-tuning training adopts the trained model on 2.4 GHz gypsum material dataset as initialization instead of random, and training parameters keep the same except cancling learning rate warm-up stage. Nevertheless, as the signal frequency increases, the difficulty of generalization also increases. To sum up, the network is able to learn the geometric distribution of VAs and the underlying physical mapping laws, demonstrating strong generalization capability.

\section{Conclusion}\label{sec5}
We proposed Mip-NeWRF, a physics-informed framework for WRF reconstruction and channel prediction that achieves high accuracy, fast convergence, and strong cross-scene robustness. Mip-NeWRF implicitly learns VA distribution from communication signals and exploits this knowledge to produce multipath level channel predictions. We introduce hybrid positional encoding for sampled intervals, adopt a MLP to predict VA probabilities and transmit amplitudes, and synthesize the receiver channel by combining network outputs with physical propagation and surface interaction attenuations. Extensive simulations show that Mip-NeWRF outperforms baseline methods with similar prediction error in larger scale scenes, and the model exhibits strong generalization across different materials and frequency bands. Future work will pursue two complementary directions. First, we will further reduce training cost by developing strategies that more rapidly focus samplings near likely VAs. Second, we will close the loop from raw received signals to channel prediction in previously unseen environments, enabling rapid, measurement-only deployment of spatial channel maps.




\bibliographystyle{IEEEtran}
\bibliography{ref_}



\vfill

\end{document}